\documentclass[a4paper, 11pt]{article}
\pdfoutput=1 

\usepackage{jheppub}

\usepackage{cleveref}
\usepackage{todonotes}
\usepackage[T1]{fontenc} 
\usepackage{subcaption}

\usepackage{pgfplots}
\pgfplotsset{compat = 1.3}
\usepgfplotslibrary{fillbetween}

\newcommand{\bulk}{\text{bulk}}
\newcommand{\brane}{\text{brane}}
\newcommand{\bdry}{\text{boundary}}
\newcommand{\total}{\text{total}}
\newcommand{\rt}{\text{RT}}

\title{\boldmath Bounds on gravitational brane couplings and tomography in AdS$_3$ black hole microstates}

\preprint{CPHT-RR039.062022}

\author[1]{Ji Hoon Lee,}
\author[1]{Dominik Neuenfeld,}
\author[1,2]{and Ashish Shukla}

\affiliation[1]{Perimeter Institute for Theoretical Physics, 31 Caroline Street N., Waterloo, Ontario N2L 2Y5, Canada}
\affiliation[2]{Centre de Physique Th\'{e}orique, CNRS, \'{E}cole Polytechnique, IP Paris, Palaiseau F-91128, France}

\emailAdd{jihoon.lee@perimeterinstitute.ca}
\emailAdd{dneuenfeld@pitp.ca}
\emailAdd{ashish.shukla@polytechnique.edu}

\abstract{We study information theoretic properties of planar black hole microstates in $2+1$ dimensional asymptotically anti-de Sitter spacetime, modeled by black holes with an end-of-the-world brane behind the horizon. 
The von Neumann entropy of sufficiently large subregions in the dual CFT exhibits a time-dependent phase, which from a doubly-holographic perspective corresponds to the appearance of quantum extremal islands in the brane description.
Considering the case where dilaton gravity is added to the brane, we show that tuning the associated couplings affects the propagation of information in the dual CFT state. By requiring that information theoretic bounds on the growth of entanglement entropy are satisfied in the dual CFT, we can place bounds on the allowed values of the couplings on the brane.
Furthermore, we initiate the study of \emph{brane tomography}, by showing how subleading corrections to the entanglement velocity can be used to learn about the properties of the brane as well as
any gravitational dynamics localized on it.}

\begin{document} 
\maketitle
\flushbottom

\section{Introduction}
\label{sec:Intro}
The concrete realization of the holographic principle \cite{tHooft:1993dmi, Susskind:1994vu} via the anti-de Sitter/con\-for\-mal field theory (AdS/CFT) correspondence \cite{Maldacena:1997re, Gubser:1998bc, Witten:1998qj} has provided us with a useful tool to probe the nature of quantum gravity in terms of the dynamics of field theories in one lower dimension and without gravity. String theory on the (warped) product of an asymptotically AdS space with a compact manifold provides exact top-down constructions of holographic duals to strongly coupled CFTs. The prototypical example is the correspondence between type IIB string theory on AdS$_5 \times$S$^5$ and $\mathcal{N} = 4$ SU($N$) super-Yang-Mills theory on the boundary. On the other hand, one can also look for effective bottom-up constructions of bulk gravitational theories in AdS, which capture certain aspects of interest for a given CFT state on the boundary.

In this paper, we explore some properties of a certain class of bottom-up models for AdS black holes dual to atypical high energy pure states in holographic CFTs. In the bulk, the states we are interested in look like planar black hole geometries which include parts of the second exterior region of the maximally extended spacetime. This second exterior region (as well as part of the black hole interior) is cut off at some finite distance. We will consider the simplest of such bottom-up constructions, where the termination of the spacetime geometry is modeled by the presence of an end-of-the-world (ETW) brane \cite{Hartman:2013qma,Takayanagi:2011zk,Fujita:2011fp,Kourkoulou:2017zaj,Almheiri:2018ijj,Cooper:2018cmb,Miyaji:2021ktr,Chandra:2022fwi}.

Apart from describing black hole microstates, these models have recently also found applications in the embedding of cosmology into holography \cite{Cooper:2018cmb, Antonini:2019qkt, Chen:2020tes, VanRaamsdonk:2020tlr, Wang:2021xih, Fallows:2022ioc, Waddell:2022fbn, Antonini:2022blk}. In cases where the brane is located at a large extrinsic curvature, there exists an effective description of the state besides the bulk and CFT perspective, as a cosmological spacetime with a thermal CFT coupled to gravity that is entangled with a CFT on a non-gravitating background \cite{randall1999alternative, Karch:2000ct}. For this case, one can think of the cosmology as being encoded in the state of the boundary CFT. 

The models we consider are also related via analytic continuation to bottom-up models of holographic BCFT states \cite{Takayanagi:2011zk, Fujita:2011fp}, see also \cite{Azeyanagi:2007qj, Cooper:2019rwk, Reeves:2021sab, Belin:2021nck, Kusuki:2021gpt, Kawamoto:2022etl, Izumi:2022opi, Anous:2022wqh}. Such constructions have been utilised for understanding the island prescription for the computation of entropies \cite{Almheiri:2019hni, Rozali:2019day,Almheiri:2019psy, Balasubramanian:2020hfs, Sully:2020pza, Geng:2020qvw, Chen:2020uac, Chen:2020hmv, Grimaldi:2022suv, Krishnan:2020fer, Deng:2020ent, May:2021zyu, Fallows:2021sge, Neuenfeld:2021wbl, Geng:2021iyq, Chu:2021gdb, Miyaji:2021lcq, Verheijden:2021yrb, Geng:2021mic, Suzuki:2022xwv, Bianchi:2022ulu, Geng:2022slq}, and also for quantum complexity \cite{Chapman:2018bqj, Ross:2019rtu, Sato:2019kik, Braccia:2019xxi, Hernandez:2020nem, Omidi:2020oit, Bhattacharya:2021jrn, Auzzi:2021ozb}.
From a top-down perspective, such an ETW brane can represent branes provided by string theory, or regions of large back-reaction where the geometry caps off \cite{DHoker:2007zhm, DHoker:2007hhe,Chiodaroli:2012vc,Bak:2020enw,Uhlemann:2021nhu,VanRaamsdonk:2021duo}. 

We will be interested in studying quantum entanglement properties of CFT states dual to such planar black holes, and their implications for the parameter space of these models. Usually, ETW branes are modeled as constant-tension branes, together with a Gibbons-Hawking-York boundary term which makes the bulk variational principle well-defined. One imposes Neumann boundary conditions for the bulk fields at the location of the brane. Since ETW branes serve as simple bottom-up models for more complicated geometries with non-trivial warping, varying fields, or large quantum-gravitational effects, it is natural---in the spirit of effective field theories---to allow for more general couplings beyond the tension term to appear in the ETW brane action. Possible terms not only include matter fields localized on the brane, such as additional scalar fields, or curvature terms of the induced metric, but also couplings to additional bulk degrees of freedom.

In this work, we focus on adding gravitational dynamics to the ETW brane action. In general dimensions one can consider the so-called Dvali-Gabadadze-Porrati (DGP) term \cite{Dvali:2000hr}, which is proportional to the Einstein-Hilbert term on the brane. Such terms can arise from the quantum effects of matter fields localized on the brane. In two dimensions, where the Einstein-Hilbert action is topological, one can have Jackiw-Teitelboim (JT) gravity \cite{JACKIW1985343,TEITELBOIM198341} on the brane. Adding such terms to the brane action changes how the Ryu-Takayanagi (RT) surfaces end on the brane, affecting their areas and hence the entanglement entropy of subregions in the dual CFT \cite{Ryu:2006bv, Ryu:2006ef, Hubeny:2007xt, Rangamani:2016dms}. This has been important in doubly-holographic models for quantum extremal islands \cite{Almheiri:2019hni, Chen:2020uac, Chen:2020hmv, Hernandez:2020nem, Grimaldi:2022suv}, where additional JT gravity terms have to be introduced on the brane in order to discuss non-trivial islands. In higher dimensions, this is not strictly necessary, but adding DGP terms makes it possible to delay or advance the onset of the island phase of extremal surfaces in models of evaporating black holes. It also enables Euclidean construction of braneworld cosmologies \cite{Fallows:2022ioc, Waddell:2022fbn}.

The possibility of adding gravitational terms to the brane action for an ETW brane behind the black hole horizon poses the question of how having such an additional term gets reflected in the properties of the dual CFT state. In this paper, we demonstrate that adding such couplings affects the growth rate of holographic entanglement entropy of subregions in the dual CFT. For concreteness, we restrict our attention to the case of AdS$_3$ planar black hole microstates with ETW branes behind their horizon. The bulk geometry then corresponds to the planar limit of the Ba\~{n}ados-Teitelboim-Zanelli (BTZ) black hole \cite{Banados:1992wn}. In addition, we allow for JT gravity localized on the two-dimensional brane. The growth rate for the entanglement entropy can be quantified in terms of the \emph{entanglement velocity} $v_E$ \cite{Hartman:2013qma}. It was conjectured in \cite{Liu:2013iza, Liu:2013qca} and proved in \cite{Hartman:2015apr} that the entanglement velocity for such states satisfies the instantaneous bound $|v_E(t)| \le 1$. For the case of eternal black holes, it is known that this inequality is obeyed at all times \cite{Hartman:2013qma}. The key observation---which will allow us to constrain the JT coupling in these models---is that for the case of single-sided black hole microstates, there will be corrections to the growth rate of entanglement entropy, coming from the fact that the areas of the RT surfaces ending on the brane are sensitive to the presence of the coupling.

We study two different types of ETW branes, distinguished by the values of their tension $T_0$. In order for our spacetime geometry to only have one asymptotic boundary (and thus to be dual to a state of a single CFT), $T_0$ must be smaller than some critical value $T_{\text{crit}}$. For this case, we obtain brane trajectories in the bulk which cut off the second asymptotic region of the maximally extended spacetime geometry, see \cref{fig:brane_trajectories_1}. On the other hand, branes with $T_0 = T_{\text{crit}}$ asymptotically approach the second asymptotic boundary, see \cref{fig:brane_trajectories_2}. This makes it less obvious that they can in fact be dual to the state of a single CFT. However, since these critical solutions can be obtained by a limiting procedure from the subcritical solutions, it is natural to consider them as well.\footnote{When $T_0 > T_{\text{crit}}$, the brane trajectory only partially cuts off the second asymptotic region, and the resulting bulk geometry includes part of the second asymptotic boundary. The holographic duals to such geometries will necessarily involve another copy of the CFT, which lives on the second asymptotic boundary. Such super-critical values for $T_0$ will not be of interest to us in the present work.}

In the subcritical case $T_0 < T_{\text{crit}}$, it turns out that depending upon the magnitude of the JT coupling added to the brane, the instantaneous bound $|v_E(t)| \le 1$ can be violated at early times. This in turn allows us to constrain the space of possible gravitational couplings on the brane. If we find that the speed bound is violated, we can conclude that the dual CFT state cannot exist and thus rule out the corresponding brane coupling. The allowed space of couplings depends on the location of the brane and is given by \cref{eq:weak_bound_2d}.

Given that there are no true singularities in AdS$_3$, one can also study extremal surfaces which continue through the singularity and end on the brane in the analytic continuation of the planar BTZ spacetime. It turns out that if we consider such surfaces, the entanglement entropy exhibits a discontinuous jump as a function of time. Thus, once again, we can rule out the corresponding coupling. This yields a much tighter bound, \cref{eq:alpha_bound}, provided one believes that such surfaces should in fact be considered.

In the case of critical tension $T_0 = T_{\text{crit}}$, there are two types of solutions related by time-reflection. Here, we find that RT surfaces can only end on the brane for a particular choice of the sign of the JT coupling. Therefore, for this case as well, we can place bounds on the JT coupling, see \cref{eq:alpha_cr_bound_weak}. Again, considering candidate RT surfaces which cross through the singularity, one can strengthen the bound, \cref{eq:alpha_cr_bound_strong}.

Additionally, the application of ETW brane models to cosmology suggests another question---how can we determine from the CFT state whether it describes a black hole with an ETW brane, and how can we extract the parameters of the brane solution? Again, focusing on planar AdS$_3$ black holes with ETW branes, in the second part of this paper we present a protocol which allows one to determine parameters of the brane solution as well as the value of the JT coupling on the brane purely from CFT data---more precisely from the late time behaviour of entanglement entropy. This constitutes a proof-of-concept for and a first step towards \emph{brane tomography}, i.e., reconstructing a brane behind the black hole horizon from the properties of the dual CFT state. 

The paper is organized as follows. In \cref{sec:setup}, we discuss ETW branes behind planar AdS$_3$ horizons and give a complete classification of translationally invariant brane solutions. This section also helps us to establish our notation for the rest of the paper and reviews some basic facts about computing entanglement entropies in the presence of ETW branes. In \cref{sec:limits_on_ve}, following \cite{Hartman:2015apr}, we present a review of the derivation of the instantaneous bound on the entanglement velocity in a translationally invariant two-dimensional CFT state with a uniform energy density. This bound provides the bedrock for our subsequent derivation of the bounds on the JT coupling on the ETW brane. In \cref{sec:subcritical_branes}, we perform a detailed study of subcritical branes, both with and without JT coupling, including the derivation of bounds on the coupling in \cref{sec:DGP_bounds_subcrit}. Subsequently, in \cref{sec:critical_branes}, the analysis is repeated for the case of critical branes. We also show how the bounds in the critical case can be obtained from the sub-critical case, providing an important cross-check for our results. \Cref{sec:tomography} lays out the protocol for performing brane tomography, where we rely on the saturation of late time entanglement growth to extract information about the brane parameters.  In \cref{sec:discussion}, we conclude the paper with a discussion and an outlook towards possible applications and extensions of the analysis presented here. The appendices provide supplementary details on various computations performed in the main text.


\section{Basic setup}
\label{sec:setup}
\subsection{Planar black holes and ETW branes}
In this paper, we consider planar black hole solutions in $2+1$-dimensional gravity with a negative cosmological constant. The gravitational action is given by
\begin{align}
\label{eq:gravity_action_vanilla}
    I_\text{gravity} = I_\bulk + I_\bdry,
\end{align}
where
\begin{align}
    I_\bulk &= \frac{1}{16\pi G_N} \int d^{3}x \sqrt{-g} \left(R + \frac{2}{L^2} \right),\label{eq:einstein_hilbert}\\
    I_\bdry &= \frac 1 {8 \pi G_N} \int d^2x \sqrt{-h} \, K + \text{(counterterms)}.\label{eq:gibb_hawk}
\end{align}
Here, $G_N$ is the $3$-dimensional bulk Newton's gravitational constant, $L$ is the AdS radius, $h_{ab}$ is the induced metric on the boundary, and $K = \nabla_\mu n^\mu$ is the extrinsic curvature of the boundary, with $n^\mu$ being the outward pointing unit normal vector. The boundary action contains terms located at the asymptotic boundary of AdS required to make the variational principle in the bulk well-defined, as well as counterterms which render the bulk on-shell action finite \cite{deHaro:2000vlm}.

A solution to the equations of motion for the above action is given by the planar limit of the Ba\~{n}ados-Teitelboim-Zanelli (BTZ) black hole geometry \cite{Banados:1992wn}. Outside the horizon at $r = r_+$, the geometry of the planar black hole is described in Schwarzschild coordinates by the metric
\begin{align}
    \label{eq:planar_metric}
    ds^2 = -f(r)dt^2 + \frac 1 {f(r)} dr^2 + \frac{r^2}{L^2} d x^2,
\end{align}
with the blackening factor
\begin{align}
    \label{eq:general_blackening}
    f(r) = \frac{r^2 - r^2_+} {L^2}.
\end{align}
The black hole has an associated temperature of $T_H = \frac{f'(r_+)}{4\pi} =\frac {r_+}{2\pi L^2}$.\footnote{Note that there is in fact no well-defined temperature or energy associated with a single planar black hole geometry. The reason is that if we rescale $t \to c t$, $r \to r/c$, and $\vec x \to c \vec x$ with some constant $c$, we change the temperature to $T_H = \frac{1}{c}\frac{r_+}{2 \pi L^2}$, while leaving the metric invariant. The black hole temperature is also the temperature of the dual CFT state. By rescaling the metric as just explained, we are essentially performing a Weyl transformation on the CFT, mapping between different states. For UV cutoff dependent quantities, this also changes the cutoff.}

As mentioned in \cref{sec:Intro}, we will be interested in planar AdS$_3$ black holes which have an end-of-the-world (ETW) brane behind their horizons.
To construct such solutions we start with an eternal planar black hole geometry, which has two exterior regions described by \cref{eq:planar_metric} connected through the black hole interior. This geometry is dual to two entangled CFTs \cite{Maldacena:2001kr}. We then introduce a time-like, co-dimension one, constant tension brane which cuts off the spacetime in the left exterior.\footnote{Choosing to cut off the left exterior is, of course, pure convention.} This removes the left asymptotic boundary from the extended spacetime, such that the resulting geometry is now dual to a state in the right CFT only, see \cref{fig:brane_trajectories}. Introducing the ETW brane modifies the total gravitational action, \cref{eq:gravity_action_vanilla}, to
\begin{align}
    \label{eq:gravity_action}
    I_\text{total} = I_\bulk + I_\brane + I_\bdry.
\end{align}
The brane action is given by
\begin{align}
\label{eq:braneaction}
    I_\brane = \frac{1}{8\pi G_N} \int d^2x \sqrt{-h} \left(K - \frac{T_0}{L} \right),
\end{align}
where $h_{ij}$ is the induced metric on the brane and $\frac{T_0}{8 \pi G_N L}$ is the constant brane tension. The normalization is chosen for later convenience. In the remainder of the paper we will simply refer to $T_0$ as the brane tension, although it should of course be understood that $T_0$ is related to the brane tension by a proportionality factor.

The equations of motion for the system are obtained by varying the action \cref{eq:gravity_action},\footnote{A careful derivation is presented in \cref{sec:brane_eom}.}
\begin{align}
    \delta I_\total =  (\text{e.o.m.}) + \frac{1}{16\pi G_N} \int_\brane d^2 x \sqrt{-h} \left(K_{ij} - K h_{ij} + \frac{T_0 }{L} \, h_{ij}\right)\delta h^{ij}.
    \label{eq:var_tot}
\end{align}
Here, we have dropped a total derivative term along the brane, and have required that the metric variation at the asymptotic AdS boundary vanishes. The term denoted by \emph{e.o.m} vanishes if the bulk equations of motion are satisfied. In order to make the second term go away, we impose Neumann boundary conditions on the brane, namely that the term which multiplies the variation $\delta h^{ij}$ vanishes. This condition can be conveniently rewritten as
\begin{align}
\label{eq:brane_eom}
K_{ij} = \frac{T_0}{L} h_{ij},
\end{align}
which places the brane at a location where the trace of its extrinsic curvature is constant. 

\subsection{Brane trajectories}
\label{sec:brane_solutions}
\begin{figure}
    \centering
    \subcaptionbox{$T_0 < T_\text{crit}$\label{fig:brane_trajectories_1}}[0.3\textwidth]{
    \begin{tikzpicture}
    \begin{scope}
    \draw[draw=none] (0,0) -- (0,4);
    \draw (0.5,0) -- (4,0) -- (4,4) -- (0.5,4);
    \draw[dashed] (0.5,0.5) -- (4,4);
    \draw[dashed] (0.5,3.5) -- (4,0);
    \draw[very thick] (0.5,0) -- (0.5,4);
    \end{scope}
\end{tikzpicture}}
    \subcaptionbox{$T_0 = T_\text{crit}$\label{fig:brane_trajectories_2}}[0.3\textwidth]{
    \begin{tikzpicture}
    \begin{axis}[x=1cm, y=1cm, xticklabels=\empty, yticklabels=\empty, xmin=-2, xmax=2, ymin=-2, ymax=2, hide axis]
        \addplot [very thick, black] file {data/brane_T1.dat};
    \end{axis}
    \begin{scope}[xshift=4cm, yshift=0cm]
    \draw (0,0) -- (0,4cm);
    \draw[dashed] (0,0) -- (-4,4);
    \draw[dashed] (0,4) -- (-2.6,1.4);
    \draw[gray, very thin] (0,0) -- (-2,0);
    \draw[gray, very thin] (0,4) -- (-4, 4);
    \end{scope}
    \end{tikzpicture}}
    \subcaptionbox{$T_0 > T_\text{crit}$\label{fig:brane_trajectories_3}}[0.3\textwidth]{
    \begin{tikzpicture}
    \begin{scope}
    \draw (0,2) -- (0,4) -- (4,4) -- (4,0) -- (0.6,0);
    \draw[dashed] (0.6,0.6) -- (4,4);
    \draw[dashed] (0,4) -- (4,0);
    \begin{axis}[x=1cm, y=1cm, xticklabels=\empty, yticklabels=\empty, xmin=-2, xmax=2, ymin=-2, ymax=2, hide axis]
        \addplot [very thick, black] file {data/brane_traj_T2.dat};
    \end{axis}
    \end{scope}
\end{tikzpicture}}
    \caption{Representative solutions for different values of the brane tension $T_0$.}
    \label{fig:brane_trajectories}
\end{figure}
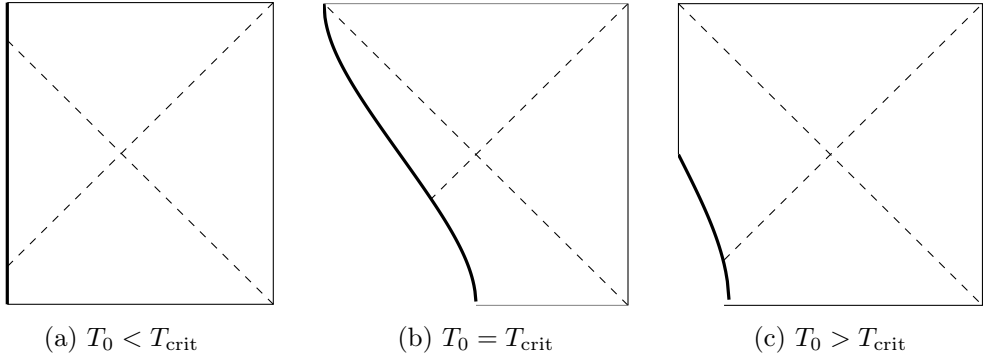

For our setup, it is natural to consider ETW branes which preserve the translation symmetry of the planar black hole geometry. Such branes follow a trajectory $r(t)$ which satisfies
\begin{align}
    \label{eq:brane_trajectory}
    \frac{dr(t)}{dt} = \pm  f(r)  \frac{L}{T_0} \sqrt{\frac{T_0^2}{L^2} - \frac{f(r)}{r^2}}.
\end{align}
To classify all possible solutions to this equation, it is convenient to choose a different parametrization for the time $t$ by defining a new coordinate $\eta$ such that
\begin{align}
    \frac{dr(\eta)}{d\eta} = \frac{dr(t)}{dt} \frac{T_0}{f(r) L}.
\end{align}
This transforms \cref{eq:brane_trajectory} into an equation describing a particle moving in a potential,
\begin{align}
    \label{eq:reparam}
    \left(\frac{dr(\eta)}{d\eta}\right)^2 + \frac{f(r)}{r^2} = \frac{T_0^2}{L^2},
\end{align}
with a fixed ``total energy'' given by $\frac{T_0^2}{L^2}$, and with the potential energy given by
\begin{align}
    V(r) =  \frac{f(r)}{r^2} = \frac 1 {L^2} \left(1 - \frac{r_+^2}{r^2}\right).
\end{align}
By examining the possible solutions $r(\eta)$ to \cref{eq:reparam}, one finds that the allowed brane trajectories can be classified into three different categories depending upon whether the tension $T_0$ is larger, smaller or equal to the critical value $T_\text{crit} = 1$. A representative brane solution for each of these categories is shown in \cref{fig:brane_trajectories}. Of course, for a fixed $T_0$, there is an infinite family of solutions related to one-another by Schwarzschild time-translations. It is also clear from the plots that the solutions with $T_0 > T_\text{crit}$ will not be of interest to us, since the left asymptotic boundary is not completely removed from the extended geometry. Nonetheless, for completeness, we include a discussion of their properties below.

\paragraph{$\mathbf{T_0 < T_\text{crit}:}$}
The first class of solutions is obtained by requiring that the brane has a turnaround point, i.e., that $\frac{dr(t)}{dt} = 0$ at some $t = t_0$. Additionally, requiring that the brane is embedded behind the horizon restricts the parameter $T_0$ to $0 \leq T_0 < T_\text{crit}$. The general solution for the brane equation of motion in Schwarzschild coordinates is then given by
\begin{align}
\label{eq:t_less_1_general_schwsch}
    r(t) = r_+ \sqrt{\frac{1 - T_0^2 \tanh^2\left(\frac{r_+ (t - t_0)}{L^2}\right)} {1 - T_0^2}}.
\end{align}

It is also interesting to consider the full trajectory in the Kruskal-Szekeres coordinates. To go to these coordinates, we use the coordinate transformation
\begin{equation}
    r = r_+ \left(\frac{1 - \tan \alpha \tan \beta}{1 + \tan \alpha \tan \beta}\right), \qquad t = \frac {L^2}{2 r_+} \log \left( - \frac {\tan \alpha} {\tan \beta}\right),
\end{equation}
which takes the metric \cref{eq:planar_metric} into the form
\begin{equation}
    ds^2 = \frac 1 {\cos^2 y} \left(- L^2 d\tau^2 + L^2 dy^2 + \frac{r_+^2}{L^2} \cos^2(\tau)\, dx^2 \right),
\end{equation}
where $\tau = \alpha + \beta$ and $y = \alpha - \beta$ run between $-\pi/2$ and $\pi / 2$. The AdS$_3$ asymptotic boundaries are located at $y = \pm \frac \pi 2$.

In Kruskal-Szekeres coordinates, the trajectory $y = y(\tau)$ of the brane is expressed most easily as a parametric equation. A nice representation is given by
\begin{align}
    \label{eq:t_less_1_general}
    \left( \cosh \frac{r_+ t_0}{L^2} \; \sin y + \sinh \frac{r_+ t_0}{L^2} \; \sin \tau \right)^2 = \frac{T_0^2}{1-T_0^2} \cos^2 y,
\end{align}
where $t_0$ is the Schwarzschild turnaround-time in the left exterior, which increases under forward time evolution. As we will see shortly, the discussion in the remainder of this paper can be made mostly independent of the choice of $t_0$. For the sake of simplicity we therefore choose the solution with $t_0 = 0$. \Cref{eq:t_less_1_general} then simplifies to
\begin{align}
    \label{eq:brane_location_vs_tension}
    \sin(y) = - T_0.
\end{align}
From this expression, it is easy to see that as $T_0$ approaches $T_\text{crit}$, the brane approaches the left asymptotic boundary at $y = - \frac{\pi}{2}$, where it stops making sense. This is of course also true in the case with generic, fixed $t_0$.\footnote{Note that the situation here is different compared to the case of spherical black holes, where in higher dimensions, the critical tension is an acceptable value for the brane tension for the turnaround solution.}
It can also be checked explicitly that the general expression, \cref{eq:t_less_1_general}, solves the equation for the brane trajectory in Kruskal-Szekeres coordinates,
\begin{align}
    \label{eq:eom_in_kruskal}
    \cos y \left( - \tan y + y'(\tau) \tan \tau \right) = T_0 \sqrt{1 - y'(\tau)^2}.
\end{align}
\Cref{fig:brane_trajectories_1} shows how the brane trajectory continues inside the black hole horizon (in the Kruskal-Szekeres coordinates). The brane emanates from the black hole singularity, exits the horizon, reaches a turning point and falls back into the black hole, eventually reaching the future singularity. The maximum separation between the brane and the black hole, denoted by $r_0$, is set by the tension through
\begin{align}
    T_0^2 = L^2 \frac{f(r_0)}{r_0^2}.
\end{align}

The induced metric on the brane cannot depend on the value of $t_0$, since Schwarzschild time shifts are an isometry of the bulk geometry. Therefore, one can deduce the induced metric on the brane by considering the special case \cref{eq:brane_location_vs_tension}. Using proper time $\lambda$ on the brane, the metric on the brane reads
\begin{align}
    \label{eq:induced_metric_T_less_1}
    ds_h^2 = - L^2 d\lambda^2 + \frac{r_+^2}{L^2} \frac{\cos^2 \left(\lambda \sqrt{1 - T^2_0}\right)}{1 - T^2_0}\,dx^2\, ,
\end{align}
which describes a big bang/big crunch cosmology with a negative cosmological constant and radiation. The proper time takes values between $\pm \frac{1}{\sqrt{1-T_0^2}} \frac \pi 2$.

\paragraph{$\mathbf{T_0 = T_\text{crit}:}$}
When the tension parameter takes on its critical value, the previous solution stops making sense, since the brane coincides with the left asymptotic AdS$_3$ boundary at $y=-\frac{\pi}{2}$. Instead, we find a new solution, where the brane emerges from the past horizon and asymptotically approaches the left boundary, as shown in \cref{fig:brane_trajectories_2}. Additionally, there is a time-reflected solution where the brane emanates from the left boundary in the asymptotic past and falls into the future horizon.

The general solution of the brane equation \cref{eq:brane_trajectory} can now be parametrized as
\begin{align}\label{eq:brane_traj_crit}
    r^2 - r_+^2 = r_+^2 e^{\frac{\pm 2 r_+ (t-t_0)}{L^2}}.
\end{align}
The sign determines whether the brane recedes from or approaches the asymptotic boundary. For the upper sign, $r(t)$ approaches the asymptotic boundary at $t \to \infty$ and the past horizon as $t \to -\infty$. Since the brane has no turnaround point anymore, the interpretation of the integration constant $t_0$ changes compared to the previous case, although Schwarzschild time translations still map between various solutions.

The meaning of $t_0$ becomes clearer if we write the brane trajectory in the Kruskal-Szekeres coordinates,
\begin{align}\label{eq:critical_family}
    \left( \sin y \pm \sin \tau \right)^2 = e^{ \mp \frac{2 r_+ t_0}{L^2}} \cos^2 y.
\end{align}
The upper sign describes a brane which hits the left boundary ($y = - \frac \pi 2$) at $\tau = \frac \pi 2$. The lower sign is the time-reflected solution. The expression shows that the brane emerges from the past (future) singularity at a location determined by the choice of $t_0$.

In fact, instead of solving the brane equation of motion, we could have arrived at the same solutions by taking \cref{eq:t_less_1_general_schwsch,eq:t_less_1_general} with $t_0 \to t_0 + \bar t$ and sending $\bar t \to \pm \infty$, while at the same time fixing
\begin{align}
\label{eq:fixed}
    (1 - T^2_0) \, e^{ \pm \frac{2 r_+ \bar t}{L^2}} = 4.
\end{align}

The induced metric on the brane describes a radiation-dominated universe with a vanishing cosmological constant,
\begin{align}
    \label{eq:induced_metric_T_eq_1}
    ds_h^2 = -\,L^2 d\lambda^2 + \frac{r_+^2}{L^2} \lambda^2 dx^2.
\end{align}
The proper time runs from $0$ to $\infty$ or $-\infty$ to $0$, depending on whether the brane emanates from he singularity in the past, or falls into it in the future, respectively.
The four-dimensional version of this solution has been discussed previously, e.g., in \cite{Gubser:1999vj}, however, with a different interpretation.

\paragraph{$\mathbf{T_0 > T_\text{crit}:}$}
In this case, the brane either emerges from the past horizon and reaches the left asymptotic boundary at a finite time, or comes out of the left asymptotic boundary at a given instant of time and falls into the future horizon. The parametrization of the brane is now given by
\begin{align}
    \label{eq:t_bigger_1_general}
    \left( \sinh \frac{r_+ t_0}{L^2} \; \sin y + \cosh \frac{r_+ t_0}{L^2} \; \sin \tau \right)^2 = \frac{T_0^2}{T_0^2-1} \cos^2 y,
\end{align}
and an example is shown in \cref{fig:brane_trajectories_3}. This expression is quite similar to \cref{eq:t_less_1_general}, except that $\sinh$ gets replaced by $\cosh$ (and vice-versa), and the sign of the right hand side gets flipped. The parameter $t_0$ gives the time at which the brane intersects the left boundary. Of course, we again have the same limiting behaviour when simultaneously taking $t_0 \to \infty$ and $T_0 \to 1$, corresponding to the marginal case $T_0 = T_\text{crit}$. 

The induced metric on the brane now models an expanding, $\lambda \in (0,\infty)$, or contracting, $\lambda \in (-\infty,0)$, spacetime,
\begin{align}
    ds_h^2 = - L^2 d\lambda^2 + \frac{r_+^2}{L^2} \frac{\sinh^2(\lambda \sqrt{T^2_0 - 1})}{T^2_0 - 1} \, dx^2,
\end{align}
with a late (early) time de-Sitter phase. Though solutions with $T_0 > T_\text{crit}$ exist formally, they do not completely remove the second asymptotic boundary of the maximally extended black hole spacetime. This makes it seem unlikely that such bulk geometries can be described holographically by a single CFT living on the right asymptotic boundary, which is why we will not discuss these solutions any further in this paper.

\subsection{The dual CFT and entanglement entropy}
Black holes in AdS spacetime have a dual description as (approximately) thermal states in a CFT, which can be thought of as living at the asymptotic boundary of the spacetime \cite{Witten:1998qj, Maldacena:2001kr}. The thermal entropy density of the CFT state agrees with the Bekenstein-Hawking entropy density of the black hole, after relating the bulk gravitational constant to the number of boundary degrees of freedom. In holographic CFTs, one can compute the von Neumann entropy for the CFT state on some subregion $A$,
\begin{align}
    S_\text{vN}(A) = - \text{tr} (\rho_A \log \rho_A),
\end{align}
with $\rho_A$ being the reduced density matrix on $A$, by using the Hubeny-Rangamani-Ryu-Takayanagi (RT) prescription \cite{Ryu:2006bv, Ryu:2006ef, Nishioka:2009un, Rangamani:2016dms}.\footnote{See \cite{Kibe:2021gtw} for a recent review of several further connections between quantum information and gravity, including bulk reconstruction and the information loss paradox.} It is given by the area of the smallest bulk extremal surface $\Sigma_\rt$ homologous to $A$,\footnote{There are subleading corrections in $\frac 1 N$, where $N$ is related to the number of degrees of freedom of the CFT, which we will ignore.} 
\begin{align}
    \label{eq:rt_formula}
    S_\text{vN}(A) = \frac{\text{Area}(\Sigma_\rt)}{4G_N}.
\end{align}

\begin{figure}
    \centering
    \subcaptionbox{Thermal surface}[0.45\textwidth]{
    \begin{tikzpicture}
        \draw (0,-2) -- (0,2);
        \draw[thick] (0,-1.5) -- (0,1.5);
        \draw[dotted] (-4,-2) -- (-4,2);
        \draw[dashed] (-2,-2) -- (-2,2);
        \draw (-3.5,-2) -- (-3.5,2);
        \node[right] (0,0) {A};
        \draw[red, thick] (0,1.5) arc (90:270:1.5);
    \end{tikzpicture}}
    \subcaptionbox{Connected surface}[0.45\textwidth]{
            \begin{tikzpicture}
        \draw (0,-2) -- (0,2);
        \draw[thick] (0,-1.5) -- (0,1.5);
        \draw[dotted] (-4,-2) -- (-4,2);
        
        \draw (-3.5,-2) -- (-3.5,2);
        \node[right] (0,0) {A};
        \draw[red, thick] (0,-1.5) -- (-3.5, -1.5);
        \draw[red, thick] (0,1.5) -- (-3.5, 1.5);
        \draw[dashed] (-2,-2) -- (-2,2);
    \end{tikzpicture}}
    \caption{The possible configurations of bulk extremal surfaces.}
    \label{fig:rt_configuration}
\end{figure}
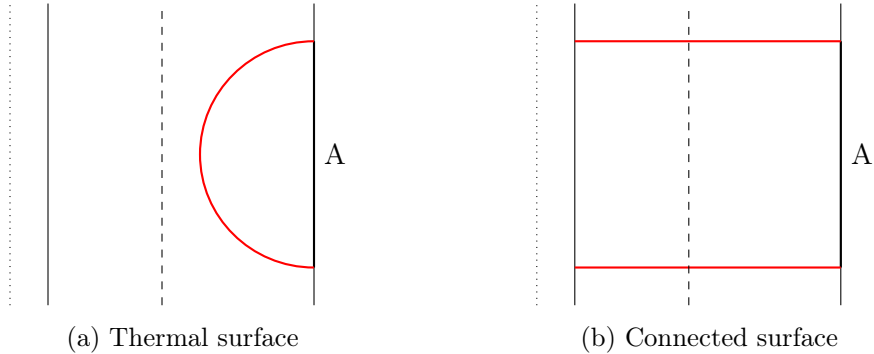

Importantly, in the presence of an ETW brane, the homology constraint only has to hold up to terms on the brane, such that the bulk RT surfaces are allowed to end on the brane \cite{Takayanagi:2011zk, Fujita:2011fp}. This becomes particularly clear in models which involve bulk defects instead of ETW branes \cite{Chen:2020uac}. In this paper we consider intervals and half-spaces in the two-dimensional CFT on the boundary. For these regions, there generally will be two possible configurations of bulk extremal surfaces. The first possibility is that the RT surface is strictly homologous to the region $A$ and stays outside the black hole horizon. Its area computes the entanglement entropy of region $A$ in the thermal state, since the geometry outside the horizon is just that of the thermofield double. We will say that this RT surface is in the \emph{thermal phase}. The area of this surface grows extensively with the size of the interval $A$. Alternatively, the RT surface can consist of two disjoint pieces which connect the boundary with the ETW brane. We will call it the \emph{connected RT surface}, or say that the RT surface is in the \emph{connected} phase. Clearly, the area of the connected surface does not depend on the size of $A$. \Cref{fig:rt_configuration} shows the two possible configurations.

The modification of the homology constraint implies that black holes with ETW branes behind their horizons are pure states, motivating their interpretation as black hole microstates. For $T_0 < T_\text{crit}$, the solution discussed above can be obtained from an analytic continuation of a Euclidean CFT construction, which shows that such states are related to regulated boundary states \cite{Kourkoulou:2017zaj, Cooper:2018cmb}. For $T_0 \geq T_\text{crit}$, as well as the cases with additional couplings discussed below, the existence of an appropriate analytic continuation to a Euclidean geometry is not clear.

\subsection{Gravitational couplings on the brane}
Bulk geometries employing ETW branes should generally be understood as effective bottom-up models which capture certain aspects of solutions to UV complete theories, such as string theory. The latter solutions will generally look much more complicated and may include, for instance, non-perturbative objects, spacetime regions in which gravity becomes strong, and warping of internal directions, to name a few. Therefore, in the spirit of constructing an effective field theory, the effective action \cref{eq:braneaction} for an ETW brane  may contain additional couplings allowed by the symmetries.

It is therefore interesting to understand to which extent are we allowed to add additional couplings to the brane. For reasons that will become clear momentarily, a particularly relevant class of couplings are gravitational couplings intrinsic to the brane. For branes with spacetime dimensionality $D \ge 3$, one can consider the Einstein-Hilbert term constructed from the induced metric on the brane as a possible additional term in the brane action. Such a term is called the Dvali-Gabadadze-Porrati term, or a \emph{DGP coupling} \cite{Dvali:2000hr}. For our setup, with two dimensional ETW branes, the Einstein-Hilbert term is purely topological. We therefore consider the simplest model of dilaton gravity instead, i.e., the Jackiw-Teitelboim (JT) model \cite{JACKIW1985343,TEITELBOIM198341}, 
\begin{align}
    \label{eq:JT_action}
    I_\text{JT} = \frac{1}{16\pi G_N^{\brane}}  \int d^2x \sqrt{-h} \, \Phi_0 R^\brane + \frac{1}{16\pi G_N^{\brane}} \int d^2x \sqrt{-h} \, \varphi \left( R^\brane - 2\Lambda^\brane \right).
\end{align}
Here $G_N^\brane$ denotes the Newton's gravitational constant on the brane. Similarly, $R^\brane$ and $\Lambda^\brane$ are the brane Ricci scalar and cosmological constant, respectively. The scalar field $\varphi$ is the \emph{dilaton}, with a constant part $\Phi_0$ that is associated with the ground state entropy.

As hinted at before, these terms have certain phenomenological features that make them particularly interesting to study. As has been discussed, e.g., in \cite{Almheiri:2019hni, Chen:2020hmv}, they affect the way RT surfaces are allowed to end on the brane. The addition of DGP couplings to the brane modifies the RT formula in \cref{eq:rt_formula} to include a \emph{contact term} at the brane\footnote{A derivation for very symmetric cases is given in the appendix of \cite{Chen:2020uac}.}
\begin{align}
    \label{eq:rt_formula_w_dgp}
    S_\text{vN}(A) = \frac{\phi(\Sigma_\rt \cap \text{brane})}{4G^\brane_N}  +  \frac{\text{Area}(\Sigma_\rt)}{4G_N}.  
\end{align}
The RT surface $\Sigma_\rt$ now has to extremize \cref{eq:rt_formula_w_dgp}. While this does not affect the bulk equations of motion for the RT surface, it does affect the location where the RT surface ends on the brane, as well as the value of the entanglement entropy (and thus the time at which transition happens between the connected and thermal extremal surfaces).

This in turn modifies the entropy if it is computed by RT surfaces which connect to the brane. In \cref{sec:subcritical_branes,sec:critical_branes} we will use this effect to place bounds on the allowed couplings in \cref{eq:JT_action}. This will be done by requiring that the growth of entropies computed via the RT prescription obeys a certain bound, which we will discuss now.


\section{Limits on entanglement velocity}
\label{sec:limits_on_ve}
In the black hole microstates of interest the entanglement entropy of sufficiently large spatial subregions in the dual CFT evolves with time. This time-evolution is subject to bounds which follow from information theoretic considerations in quantum field theory. In the present section, we discuss bounds on the \emph{entanglement velocity} in two-dimensional CFTs, which will be useful for the analysis in \cref{sec:subcritical_branes,sec:critical_branes} for imposing restrictions on the space of allowed gravitational couplings on the ETW branes. In \cref{sec:discussion}, we briefly comment upon information theoretic bounds in higher dimensions.

Entanglement velocity is a useful measure of the instantaneous rate of entanglement growth in a translation-invariant CFT state with a uniform energy density. For a spatial subregion $A$ in such a CFT state, with boundary $\partial A$ and entanglement entropy $S(A)$, the entanglement velocity is defined via
\begin{align}
    \label{eq:def_ve}
    v_E \equiv \frac{\partial_t S(A)}{s_\text{eq} |\partial A|},
\end{align}
where $|\partial A|$ is the volume of $\partial A$, and $s_{\text{eq}}$ denotes the entropy density for the system if it were in a state of thermal equilibrium with the same energy-density as the CFT state of interest. It is important to note that despite the name the entanglement velocity is not a physical velocity and thus there is no a priori reason for it to be bounded from above by the speed of light.

Usually, $v_E$ is discussed in the context of late time entanglement growth for quenched quantum systems. By late times one means time scales much larger than the inverse temperature $\beta$, but smaller than the time at which the entropy growth saturates. 
However, for our purposes, we are interested in $v_E$ and local bounds on its value which also hold at early times, i.e., at times $t \lesssim \beta$. 
The authors of \cite{Liu:2013iza, Liu:2013qca} conjectured an early time bound on $v_E$ given by the speed of light in any number of dimensions.
For the case of two dimensional quantum field theories, an argument constraining the instantaneous value of $v_E$ to $v_E \le 1$ at all times appeared in \cite{Hartman:2015apr}. Given the reliance of our subsequent discussion on this constraint, we now provide a quick review of the main ideas involved in the derivation of \cite{Hartman:2015apr}, suitably modified for our purposes.

We denote by $\rho$ the density matrix for the pure CFT state of interest, modeled in the bulk by a black hole microstate with an ETW brane behind the horizon. The state is translationally invariant and has a uniform energy density. Also, $\rho^{(\beta)}$ denotes the density matrix for the thermal state of the CFT at an inverse temperature $\beta$, chosen such that the thermal state has the same energy density as $\rho$. Now, for a connected subregion $A$ in the boundary CFT, we consider the relative entropy between the state on $A$ and the thermal state,
\begin{equation}
\begin{split}
    \label{eq:rel_entropy}
    S(\rho_A||\rho_A^{(\beta)}) &\equiv \text{tr} (\rho_A \log \rho_A) - \text{tr}(\rho_A \log \rho_A^{(\beta)}) \\
    &= S(\rho_A^{(\beta)}) - S(\rho_A) + \langle K^{(\beta)}_A \rangle - \langle K^{(\beta)}_A \rangle_{\beta},
\end{split}
\end{equation}
where $\rho_A \equiv \text{tr}_{\bar A}\rho$ is the reduced density matrix on $A$ in the state $\rho$, and $\rho^{(\beta)}_A \equiv \text{tr}_{\bar A} \rho^{(\beta)}$ is the reduced density matrix on $A$ in the thermal state.
Furthermore, $K^{(\beta)}_A$ is the modular Hamiltonian associated to the subregion $A$ in the thermal state, and is defined via $\rho^{(\beta)}_A = e^{-K_A^{(\beta)}}/\text{tr}(e^{-K_A^{(\beta)}})$. Also, $\langle\ldots\rangle \equiv \text{tr}(\rho\ldots)$ and $\langle\ldots\rangle_\beta \equiv \text{tr}(\rho^{(\beta)}\ldots)$. In a local and relativistic quantum field theory, where regions $A$ and $B$ have domains of dependence $\mathcal D(A)$ and $\mathcal D(B)$, respectively, we have that
\begin{align}
    \label{eq:rel_entropy_ineq}
    S(\rho_A||\rho_A^{(\beta)}) \leq S(\rho_B||\rho_B^{(\beta)}) \qquad \text{ for } \qquad \mathcal D(A) \subset \mathcal D(B),
\end{align}
the reason being that relative entropy cannot increase under a partial trace, a property known as monotonicity of relative entropy. These properties of relative entropy, together with the properties of the state $\rho$ we are considering, yield an immediate upper bound on the growth of entanglement entropy, as we now discuss.
\begin{figure}
    \centering
    \begin{tikzpicture}
    \draw [gray, thin] (0,0) -- (2,2) -- (0,4) -- (-2,2) -- cycle;
    \draw [black, thick] (-2,2) -- node[above] {B} (2,2);
    \draw [gray, thin] (0,0) -- (1.5,1.5) -- (0,3) -- (-1.5,1.5) -- cycle;
    \draw [black, thick] (-1.5,1.5) -- node[below] {A} (1.5,1.5);
    
    \draw [<->] (1.5,0) -- node[below] {$\Delta x = \Delta t$} (2,0);
    \draw [gray, dotted] (1.5, 0) -- (1.5,1.5);
    \draw [gray, dotted] (2, 0) -- (2,2);
    
    \draw [<->] (2.5,1.5) -- node[right] {$\Delta t$} (2.5,2);
    \draw [gray, dotted] (1.5, 1.5) -- (2.5,1.5);
    \draw [gray, dotted] (2, 2) -- (2.5,2);
    
    \end{tikzpicture}
    \caption{Two regions $A$ and $B$. Region $B$ is larger than $A$ by $2 \Delta t$ and is located $\Delta t$ to the future of $A$.}
    \label{fig:domains_of_dependence}
\end{figure}
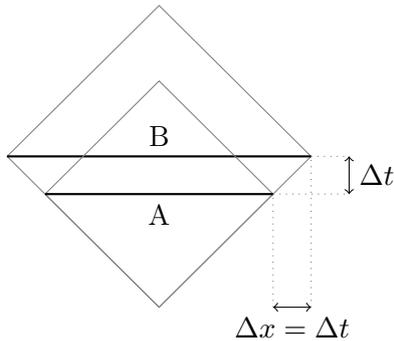

In a two-dimensional CFT the thermal modular Hamiltonian of an interval $A = [a, b]$ is a local integral of the stress-energy tensor \cite{Cardy:2016fqc},
\begin{align}
    K_A^{(\beta)} = \frac \beta {2 \pi} \int_a^b dx \, \frac{\left(1 - e^{- \frac{2 \pi (x-a)}{\beta}}\right)\left(1 - e^{- \frac{2 \pi (b-x)}{\beta}}\right)}{\left(1 - e^{- \frac{2 \pi (b-a)}{\beta}}\right)}\, T_{00}(x-a).
\end{align}
In our case, the expectation value of the stress-energy tensor is the same in the thermal state as well as the microstate, $\langle T_{00}\rangle_\beta = \langle T_{00} \rangle$, since we chose our reference thermal state to have the same energy density as the microstate. From a bulk perspective, this happens because the geometry outside the horizon is identical for both the cases. The last two terms in \cref{eq:rel_entropy} therefore cancel and we are left with
\begin{align}
    S(\rho_A||\rho_A^{(\beta)}) = S(\rho_A^{(\beta)}) - S(\rho_A).
    \label{eq:simp_rel}
\end{align}

We now choose another interval $B$ of length $|B| = |A| + 2\Delta t$ and located $\Delta t$ to the future of $A$, see \cref{fig:domains_of_dependence}. Clearly, for this choice $\mathcal D(A) \subset \mathcal D(B)$. We can therefore make use of the inequality \cref{eq:rel_entropy_ineq} along with \cref{eq:simp_rel} and conclude that
\begin{align}
    \label{eq:inequalities_entropy}
    S(\rho_B) - S(\rho_A) \leq S(\rho^{(\beta)}_B) - S(\rho^{(\beta)}_A).
\end{align}
The entanglement entropies in the thermal state, $S(\rho^{(\beta)}_{A})$ and $S(\rho^{(\beta)}_{B})$, are time-independent. Moreover, it follows from translation-invariance that the extremal surfaces which compute the entanglement entropy in the connected phase, i.e., when the interior surface give the entropy, fall straight into the black hole and connect to the ETW brane. As a result, and at leading order in $\frac 1 N$, the entanglement entropies $S(\rho_{A}), S(\rho_{B})$ do not depend upon the width of the interval $A$ or $B$, as long as we are in the connected phase. Assuming that the entropy of $\rho_{A}$, $\rho_B$ are indeed computed using extremal surfaces in the connected phase, we can divide \cref{eq:inequalities_entropy} by $\Delta t$ and take the limit $\Delta t \to 0$ to obtain
\begin{align}
    2 \partial_\ell S(\rho^{(\beta)}_A) \geq \partial_t S(\rho_A) ,
\end{align}
where $\ell = |A|$ is the length of the interval $A$. Using the explicit formula for the thermal entropy of a two-dimensional CFT for an interval of length $|A|$ we obtain
\begin{align}
    \label{eq:inequality_intermediate}
    \frac{r_+}{2 G_N L} \coth\left(\frac{r_+ |A|}{2 L^2}\right) \geq \partial_t S(\rho_A).
\end{align}
The equilibrium entropy density for our setup is given by $s_\text{eq} = {r_+}/{4 G_N L}$, and the inverse temperature by $\beta = 2\pi L^2/r_+$. We can use these expressions together with $|\partial A| = 2$ to rewrite \cref{eq:inequality_intermediate} as
\begin{align}
    \label{eq:future_bound}
    \frac{\partial_t S(\rho_A)}{s_\text{eq} |\partial A|} \leq \coth\left(\frac{\pi |A|}{\beta}\right).
\end{align}

The argument presented above can be repeated by choosing a region $B'$ which is in the past of $A$ by an amount $\Delta t$, and is $\Delta x = 2 \Delta t$ larger than $A$. This gives the opposite sign on the left hand side of \cref{eq:future_bound}. The two bounds can be succinctly summarized as
\begin{align}
    \label{eq:ve_bound_2d_exact}
    \left| v_E(t) \right| \leq \coth\left(\frac{\pi |A|}{\beta}\right).
\end{align}
The crucial point to note here is that nowhere in the argument did we refer to any particular instant of time, and so this bound on $v_E$ holds at all times. We can further assume that the region size is much larger than the inverse temperature scale, $|A| \gg \beta$, to get
\begin{align}
    \label{eq:ve_bound_2d}
    \left| v_E(t) \right| \leq 1.
\end{align}
Since we are considering a planar black hole, we can always achieve the limit $|A| \gg \beta$. Moreover, this is precisely the limit we are interested in, since it is in this limit that $S(\rho_A)$ is computed by an extremal surface in the connected phase.


\section{Subcritical branes}
\label{sec:subcritical_branes}

\begin{figure}[t]
    \centering
    \subcaptionbox{}[0.45\textwidth]{
    \begin{tikzpicture}
    \draw (0,0) node [below] {$y = y_\bdry$} -- (0,4);
    \draw[dashed] (0,0) -- (-3.5,3.5);
    \draw[dashed] (0,4) -- (-3.5,0.5);
    \draw[->] (0.5,1.5) -- node [right] {$\tau$}(0.5,2.5);
    \draw[very thick] (-3.5,0) node [below] {$y = y_\brane$} -- (-3.5,4);
    \draw[gray, very thin] (0,0) -- (-3.5,0);
    \draw[gray, very thin] (0,4) -- (-3.5, 4);
    \draw[->] (-2.5, 4.5) -- node[above] {$y$} (-1.5, 4.5);
    \end{tikzpicture}}
    \subcaptionbox{}[0.45\textwidth]{
    \begin{tikzpicture}
    \draw (0,0) -- (0,4);
    \draw[dashed] (0,0) -- (-3.5,3.5);
    \draw[dashed] (0,4) -- (-3.5,0.5);
    \draw[very thick] (-3.5,0) node [below] {\phantom{$y = y_\brane$}} -- (-3.5,4);
    \draw[gray, very thin] (0,0) -- (-3.5,0);
    \draw[gray, very thin] (0,4) -- (-3.5, 4);
    \draw[gray, thin] (0,3.91) -- (-3.5,3.91);
    \draw[red, thick] (0,3.52) -- (-3.5,3.52);
    \draw[gray, thin] (0,3.15) -- (-3.5,3.15);
    \draw[gray, thin] (0,2.76) -- (-3.5,2.76);
    \draw[gray, thin] (0,2.38) -- (-3.5,2.38);
    \draw[red, thick] (0,2) -- (-3.5,2);
    \draw[gray, thin] (0,1.62) -- (-3.5,1.62);
    \draw[gray, thin] (0,1.24) -- (-3.5,1.24);
    \draw[gray, thin] (0,0.85) -- (-3.5,0.85);
    \draw[red, thick] (0,0.47) -- (-3.5,0.47);
    \draw[gray, thin] (0,0.09) -- (-3.5,0.09);    
    \draw[->] (0.5,1.5) -- node [above, rotate = -90] {$\tau_\bdry$}(0.5,2.5);
    \draw[->] (-4,1.5) -- node [above, rotate = 90] {$\tau_\brane$}(-4,2.5);
    \end{tikzpicture}}
    \caption{The Penrose diagram for an AdS$_3$ black hole microstate with an embedded ETW brane. (a) The asymptotic boundary is located on the right. The ETW brane cuts off the geometry on the left. Horizons are represented by dashed lines. (b) The horizontal lines are extremal surfaces in the connected phase.}
    \label{fig:two_dimensions_vanilla}
\end{figure}
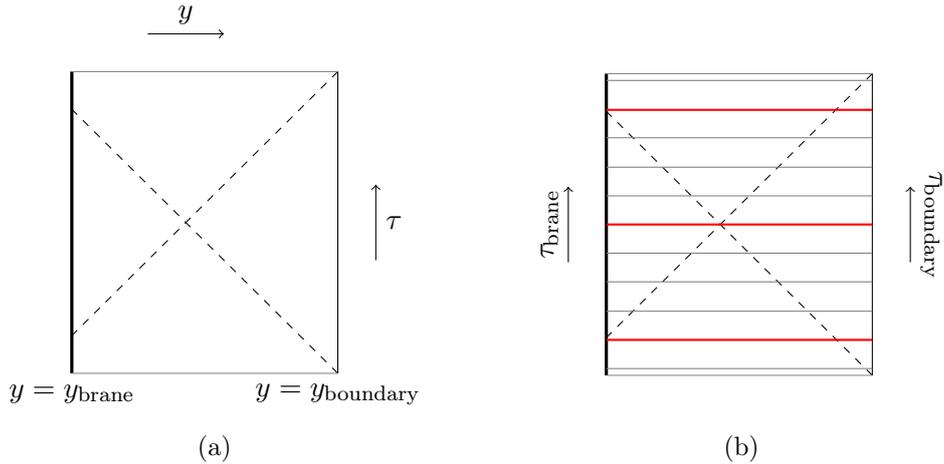

We now have all the ingredients to compute the entanglement velocity in the microstate geometries introduced in \cref{sec:setup}. In this section, we will consider branes with $T_0 < T_\text{crit}$. Branes at critical tension, $T_0 = T_\text{crit}$, will be discussed in the next section.

\subsection{Branes with subcritical tension}
\label{sec:subcr_vanishing_coupling}
In Kruskal-Szekeres coordinates, the region of spacetime which would have been close to $y = - \frac \pi 2$ is cut off by the ETW brane. Due to the translation symmetry in the transverse direction, the brane location is completely specified by the profile $y_\brane(\tau) = y(\tau)$. As seen in the previous section, the entanglement velocity is proportional to the time-derivative of the entanglement entropy. The range of values it can take on is therefore insensitive to time translations. We will thus focus on the solution \cref{eq:brane_location_vs_tension}, which is a special case of \cref{eq:t_less_1_general} with $t_0 = 0$.

In order to compute the entanglement velocity, we need to know the entanglement entropy $S(A)$ for an interval of length $|A|$ at the boundary. Of course, this is done by computing the area of the correct Ryu-Takayanagi (RT) surface homologous to the boundary interval. Areas for extremal surfaces anchored on the right asymptotic boundary diverge and therefore need to be regulated by introducing a radial cutoff. In Schwarzschild coordinates, the cutoff is chosen at $r_\epsilon = \frac {L} \epsilon$, with $\epsilon \ll 1$. In Kruskal-Szekeres coordinates $(\tau, y)$ this translates to a time-dependent cutoff surface,
\begin{align}\label{eq:def_cutoff_y}
    y_\epsilon = \frac \pi 2 - \frac{r_+ \epsilon}{L} \cos \tau + \mathcal O(\epsilon^2).
\end{align}

For the situation at hand, we have two candidate bulk extremal surfaces, as shown in \cref{fig:rt_configuration}. The correct entanglement entropy is determined by choosing the extremal surface with the smaller area. Both extremal surfaces and their areas can be computed via the usual methods. Given an interval of size $|A|$ at the boundary and some time $\tau$, we are looking for an extremal surface in the bulk, which is homologous to the boundary interval. The thermal surface, which stays outside the horizon, is located at a constant Schwarzschild time $t$. The area associated with this surface gives the well-known result for the entropy of an interval in a two-dimensional CFT at a finite temperature, which when expressed in terms of the bulk quantities reads
\begin{align}
    \label{eq:exterior_entropy_2d}
    S^{\text{(th)}}_\text{vN}(A) = \frac{L}{2G_N} \log\left(\frac{2L}{r_+ \epsilon}\,  \sinh \left(\frac{r_+ |A|}{2 L^2}\right)\right).
\end{align}

On the other hand, the connected extremal surface consists of two components, which are located at constant values of $x$, corresponding to the two end-points of the boundary interval $A$. The components shoot out into the bulk and terminate on the ETW brane. We can therefore parametrize the trajectory of this connected extremal surface $\Sigma_{\text{conn}}$ by $\tau(y)$. The associated area functional
\begin{align}
    \label{eq:area_functional_2d}
    A(\Sigma_\text{conn}) = 2 L \int^{y_\epsilon}_{y_{\text{brane}}} dy \, \frac{\sqrt{1 - \tau'(y)^2}}{\cos y}
\end{align}
only depends on the derivative $\partial_y \tau(y) \equiv \tau'(y)$, but is independent of $\tau(y)$. This gives rise to the conserved quantity
\begin{align}
    \label{eq:conserved_charge_2d}
    \mathcal Q_{E} = - \frac{\delta}{\delta \tau'(y)} \frac{A(\Sigma_\text{conn})}{2L} = \frac{\tau'(y)}{\cos(y) \sqrt{1 - \tau'(y)^2}}.
\end{align}
This equation can be solved for $\tau'(y)$ and integrated to yield an analytic solution for the trajectory of the connected extremal surface,
\begin{align}
    \label{eq:rt_trajectory_2d}
   \tau(y) = \tau_\bdry + \arcsin \left(\frac{\mathcal Q_E}{\sqrt{1 + \mathcal Q_E^2}} \sin y \right) - \arcsin\left(\frac{\mathcal Q_E}{\sqrt{1 + \mathcal Q_E^2}}\right).
\end{align}
Substituting this into the area functional \cref{eq:area_functional_2d} yields 
\begin{align}
    \label{eq:area_connected}
    A(\Sigma_\text{conn})  = 2L\, \text{arcsinh} \left. \left(\frac{\tan y}{\sqrt{1 + \mathcal Q_E^2}}\right) \right|^{y_{\epsilon}}_{y_{\brane}}.
\end{align}

So far our computations have given us the most general solution for the connected extremal surface dictated by the bulk geometry alone. However, we have not yet imposed the boundary condition for the extremal surface at the location where it intersects the brane. This is closely connected with the yet-to-be-determined conserved charge $\mathcal Q_E$. Upon varying \cref{eq:area_functional_2d} with respect to $\tau(y)$, we obtain two boundary terms
\begin{align}
    \label{eq:bounday_cond_2d}
   0 =  \frac{2L}{\cos y} \frac{\tau'(y)}{\sqrt{1 - \tau'(y)^2}} \delta \tau(y) \bigg|^{y_\epsilon}_{y_\brane} = 2 L \mathcal Q_{E} \, \delta \tau(y) \bigg|^{y_\epsilon}_{y_\brane}.
\end{align}
At the asymptotic boundary, $\delta \tau(y)$ vanishes, since the extremal surface is fixed there. On the other hand, at the location of the brane the extremal surface can move freely, so that we can impose the Neumann boundary condition, and $\delta\tau(y)|_{y_{\text{brane}}} \neq 0$. Thus, for \cref{eq:bounday_cond_2d} to be satisfied, $\mathcal Q_E$ must vanish. In conclusion, we find that the entropy $S^{(\text{conn})}_\text{vN}(A) = A(\Sigma_{\text{conn}})/4G_N$ associated with the connected extremal surface $\Sigma_{\text{conn}}$ is
\begin{align}
    \label{eq:interior_entropy_2d}
    \begin{split}
    S^{(\text{conn})}_\text{vN}(A) &= \frac L {2 G_N} \left[ \text{arcsinh} \left( \tan \left( \frac \pi 2 - \frac{r_+\epsilon}L \cos \tau \right)\right) - \text{arcsinh} \left( \tan y_\brane \right) \right]\\
    &= \frac L {2 G_N} \log\left(\frac{2L}{r_+\epsilon \cos(\tau)} \sqrt{\frac{1+T_0}{1 - T_0}}\right) + \mathcal O(\epsilon^2).
    \end{split}
\end{align}
The unusual factor of two comes from the presence of two components of $\Sigma_{\text{conn}}$.
Obviously, this entropy is independent of $|A|$. By going to the large area limit $|A| \to \infty$, we can make sure that the extremal surface in the connected phase has the smaller area, and hence the entanglement entropy up to arbitrarily late times is given by \cref{eq:interior_entropy_2d}. In order to compute the entanglement velocity, \cref{eq:def_ve}, we need to make sure to take the derivative with respect to the Schwarzschild time. The result,
\begin{align}
    |v_E| = \left|\tanh\left(\frac{r_+ t}{L^2}\right)\right| < 1,
\end{align}
evidently obeys the bound \cref{eq:ve_bound_2d} for all times.

\subsection{Introducing JT gravity on the brane}
\label{sec:DGP2D}
Let us now investigate the situation with additional gravitational dynamics localized on the ETW brane. We choose to add Jackiw-Teitelboim gravity to the brane \cite{TEITELBOIM198341, JACKIW1985343}, since pure Einstein gravity is topological in two dimensions. More precisely, we will set $T_0 = 0$ and augment the total action, \cref{eq:gravity_action}, with the JT action, \cref{eq:JT_action}.\footnote{It is also possible to add a counterterm $I_\text{ct} = - \frac{1}{4\pi G_N}\int \sqrt{h}$ to the action \cite{Balasubramanian:1999re}, such that the dilaton couples canonically to the stress-energy tensor, c.f., the discussion in \cite{Chen:2020uac}. However, since this will not change the arguments below, we refrain from doing so.} In the discussion above, $T_0$ controlled the location of the brane. Now this role is played by $\Lambda^\brane$. The dilaton equation of motion imposes the condition $R^\brane = 2\Lambda^\brane$ which determines the location of the brane through
\begin{align}
    - \frac{\cos^2 y_\brane}{L^2} = \Lambda^\brane.
\end{align}
As before, the brane is located at a constant $y = y_\brane$, and the induced metric on the brane is given by \cref{eq:induced_metric_T_less_1}. Now, however, we should write $\cos(y_\brane)$ instead of $\sqrt{1 - T_0^2}$. The metric equation of motion can be brought into the form
\begin{align}
    \label{eq:eom_dilaton}
    \nabla_i \nabla_j \varphi + \Lambda^\brane \varphi h_{ij} = \frac{G_N^\brane}{G_N} K_{ij}.
\end{align}
The inhomogeneous part of the equation above can be solved by a constant contribution $\varphi_0$ to $\varphi$, given by
\begin{align}
    \label{eq:inhomo}
    \varphi_0 = \frac{G_N^\brane}{2 G_N \Lambda^\brane} K.
\end{align}
Including the homogeneous solution to \cref{eq:eom_dilaton}, the full solution for the dilaton is then given by
\begin{align}
    \label{eq:dilaton_def}
    \varphi(\tau_\brane) = \varphi_0 + \varphi_1 \sin \tau_\brane.
\end{align}
This solution is the dilaton profile between the inner and outer horizons of a two-sided black hole in JT gravity.
Here, $\varphi_0$ is determined by \cref{eq:inhomo} and $\varphi_1$ is a free parameter, which we will try to constrain below.
For later reference, let us also note that the solution expressed in proper time reads
\begin{align}
    \label{eq:subcr_dil_proper_time}
    \varphi(\tau_\brane) = \varphi_0 + \varphi_1 \sin \left(\lambda \sqrt{- \Lambda^\brane L} \right).
\end{align}

Let us now turn to the computation of the extremal surfaces. Since the thermal extremal surface stays outside the black hole horizon, it is clear that it is not affected by adding JT gravity to the brane, which sits behind the horizon. Consequently, \cref{eq:exterior_entropy_2d} for the entropy associated with the thermal extremal surface homologous to a boundary interval of size $|A|$ on a constant-$\tau$ slice still remains valid.

In contrast, the connected extremal surfaces end on the brane and are sensitive to the presence of additional gravitational dynamics on the brane. The functional which needs to be extremized as well as evaluated in order to obtain the correct boundary entropy associated with the connected extremal surface depends upon the value of the dilaton, $\Phi = \Phi_0 + \varphi$, at the intersection of the extremal surface with the brane. It is given by \cref{eq:rt_formula_w_dgp}, which in our case reads
\begin{align}
    \label{eq:s_dgp_2d}
    S^\text{(conn)}_{\text{vN}}(\Phi) = \frac{\Phi_0 + \varphi(\tau_{\brane})}{2 G_N^{\brane}} + \frac{L}{2G_N} \int_{y_{\brane}}^{y_{\epsilon}} dy \, \frac{\sqrt{1-\tau'(y)^2}}{\cos y},
\end{align}
evaluated on the smallest extremum. Since we have only modified the area functional at the location of the brane, the general analysis of \cref{sec:subcr_vanishing_coupling}, and in particular \cref{eq:rt_trajectory_2d} carries over. However, the discussion we presented surrounding \cref{eq:bounday_cond_2d}, where we argued that the conserved charge $\mathcal Q_E$ associated with the connected extremal surface needs to vanish, has to be revisited. Due to the presence of the first term in \cref{eq:s_dgp_2d}, \cref{eq:bounday_cond_2d} obtains a new contribution and now reads
\begin{align}
    \mathcal Q_E = -\frac{G_N}{G_N^{\brane} L}\, \varphi'(\tau_{\brane}) = -\frac{G_N \varphi_1}{G_N^{\brane}L}\cos(\tau_{\brane}),
    \label{eq:extremcond}
\end{align}
where $\varphi'$ is the derivative of $\varphi$ with respect to $\tau_\brane$. The quantity $\varphi_1$ was defined in \cref{eq:dilaton_def}. Thus, the value of the conserved charge $\mathcal Q_E$ at any given time depends on the derivative of the dilaton at the location where the extremal surface intersects the brane. For later convenience, we introduce the notation
\begin{align}
    \label{eq:def_alpha}
    \alpha \equiv \frac{G_N \varphi_1}{G_N^{\brane} L}, 
\end{align}
which will subsequently be termed as the ``JT coupling,'' such that 
\begin{equation}
    \mathcal Q_E= - \alpha \cos(\tau_{\brane}) = -\alpha \cos\left(\lambda\sqrt{-\Lambda^\brane L}\right).
    \label{eq:QEalpha}
\end{equation}

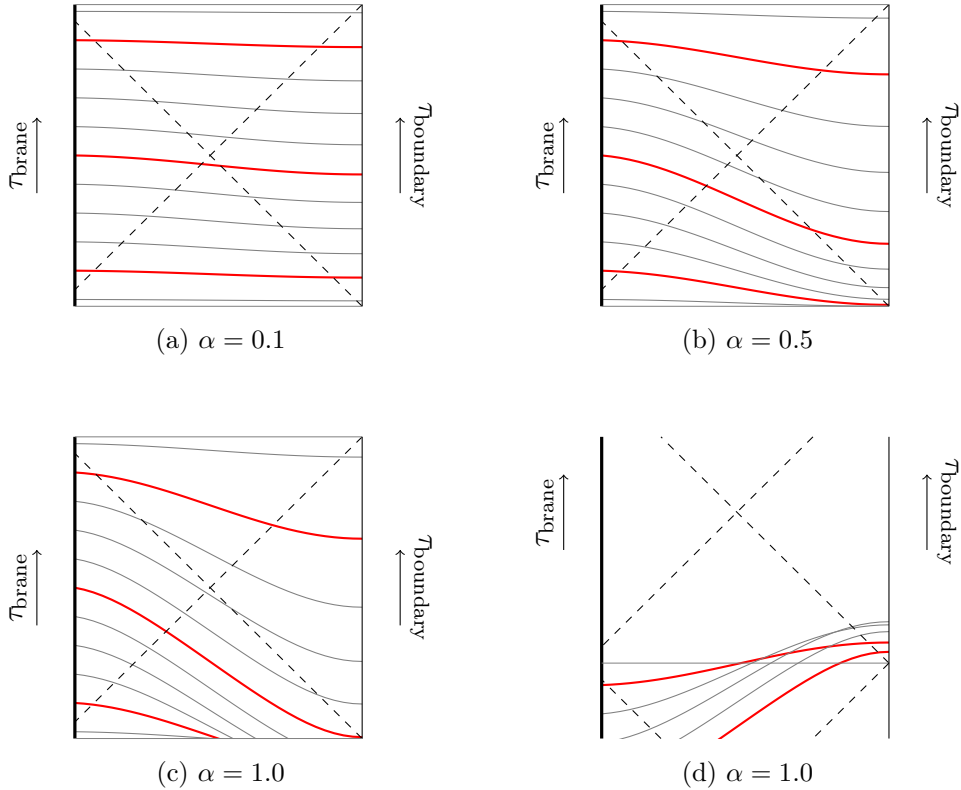
\begin{figure}
    \centering
    \subcaptionbox{$\alpha = 0.1$}[0.45\textwidth]{
    \begin{tikzpicture}
    \begin{axis}[x=1cm, y=1cm, xticklabels=\empty, yticklabels=\empty, xmin=-2, xmax=2, ymin=-2, ymax=2, hide axis]
        \addplot [thick, red] file {data/rt_surface_t+0.0_phi0.1_ybrn1.4.dat};
        \addplot [thin, gray] file {data/rt_surface_t+0.3_phi0.1_ybrn1.4.dat};
        \addplot [thin, gray] file {data/rt_surface_t+0.6_phi0.1_ybrn1.4.dat};
        \addplot [thin, gray] file {data/rt_surface_t+0.9_phi0.1_ybrn1.4.dat};
        \addplot [thick, red] file {data/rt_surface_t+1.2_phi0.1_ybrn1.4.dat};
        \addplot [thin, gray] file {data/rt_surface_t+1.5_phi0.1_ybrn1.4.dat};
        \addplot [thin, gray] file {data/rt_surface_t-0.3_phi0.1_ybrn1.4.dat};
        \addplot [thin, gray] file {data/rt_surface_t-0.6_phi0.1_ybrn1.4.dat};
        \addplot [thin, gray] file {data/rt_surface_t-0.9_phi0.1_ybrn1.4.dat};
        \addplot [thick, red] file {data/rt_surface_t-1.2_phi0.1_ybrn1.4.dat};
        \addplot [thin, gray] file {data/rt_surface_t-1.5_phi0.1_ybrn1.4.dat};
    \end{axis}
    \begin{scope}[xshift=4cm, yshift=0cm]
    \draw (0,0) -- (0,4cm);
    \draw[dashed] (0,0) -- (-3.78,3.78);
    \draw[dashed] (0,4) -- (-3.78,0.22);
    \draw[very thick] (-3.78,0)  -- (-3.78,4);
    \draw[gray, very thin] (0,0) -- (-3.78,0);
    \draw[gray, very thin] (0,4) -- (-3.78, 4);
    \draw[->] (0.5,1.5) -- node [above, rotate=-90] {$\tau_\bdry$}(0.5,2.5);
    \draw[->] (-4.28,1.5) -- node [above, rotate=90] {$\tau_\brane$}(-4.28,2.5);
    \end{scope}
    \end{tikzpicture}}
    \subcaptionbox{$\alpha = 0.5$}[0.45\textwidth]{
    \begin{tikzpicture}
    \begin{axis}[x=1cm, y=1cm, xticklabels=\empty, yticklabels=\empty, xmin=-2, xmax=2, ymin=-2, ymax=2, hide axis]
        \addplot [thick, red] file {data/rt_surface_t+0.0_phi0.5_ybrn1.4.dat};
        \addplot [thin, gray] file {data/rt_surface_t+0.3_phi0.5_ybrn1.4.dat};
        \addplot [thin, gray] file {data/rt_surface_t+0.6_phi0.5_ybrn1.4.dat};
        \addplot [thin, gray] file {data/rt_surface_t+0.9_phi0.5_ybrn1.4.dat};
        \addplot [thick, red] file {data/rt_surface_t+1.2_phi0.5_ybrn1.4.dat};
        \addplot [thin, gray] file {data/rt_surface_t+1.5_phi0.5_ybrn1.4.dat};
        \addplot [thin, gray] file {data/rt_surface_t-0.3_phi0.5_ybrn1.4.dat};
        \addplot [thin, gray] file {data/rt_surface_t-0.6_phi0.5_ybrn1.4.dat};
        \addplot [thin, gray] file {data/rt_surface_t-0.9_phi0.5_ybrn1.4.dat};
        \addplot [thick, red] file {data/rt_surface_t-1.2_phi0.5_ybrn1.4.dat};
        \addplot [thin, gray] file {data/rt_surface_t-1.5_phi0.5_ybrn1.4.dat};
    \end{axis}
    \begin{scope}[xshift=4cm, yshift=0cm]
    \draw (0,0) -- (0,4cm);
    \draw[dashed] (0,0) -- (-3.78,3.78);
    \draw[dashed] (0,4) -- (-3.78,0.22);
    \draw[very thick] (-3.78,0)  -- (-3.78,4);
    \draw[gray, very thin] (0,0) -- (-3.78,0);
    \draw[gray, very thin] (0,4) -- (-3.78, 4);
    \draw[->] (0.5,1.5) -- node [above, rotate=-90] {$\tau_\bdry$}(0.5,2.5);
    \draw[->] (-4.28,1.5) -- node [above, rotate=90] {$\tau_\brane$}(-4.28,2.5);
    \end{scope}
    \end{tikzpicture}} 
    \subcaptionbox{$\alpha = 1.0$\label{fig:rt_phi1}}[0.45\textwidth]{
    \begin{tikzpicture}
    \begin{axis}[x=1cm, y=1cm, xticklabels=\empty, yticklabels=\empty, xmin=-2, xmax=2, ymin=-2, ymax=2, hide axis]
        \addplot [thick, red] file {data/rt_surface_t+0.0_phi1.0_ybrn1.4.dat};
        \addplot [thin, gray] file {data/rt_surface_t+0.3_phi1.0_ybrn1.4.dat};
        \addplot [thin, gray] file {data/rt_surface_t+0.6_phi1.0_ybrn1.4.dat};
        \addplot [thin, gray] file {data/rt_surface_t+0.9_phi1.0_ybrn1.4.dat};
        \addplot [thick, red] file {data/rt_surface_t+1.2_phi1.0_ybrn1.4.dat};
        \addplot [thin, gray] file {data/rt_surface_t+1.5_phi1.0_ybrn1.4.dat};
        \addplot [thin, gray] file {data/rt_surface_t-0.3_phi1.0_ybrn1.4.dat};
        \addplot [thin, gray] file {data/rt_surface_t-0.6_phi1.0_ybrn1.4.dat};
        \addplot [thin, gray] file {data/rt_surface_t-0.9_phi1.0_ybrn1.4.dat};
        \addplot [thick, red] file {data/rt_surface_t-1.2_phi1.0_ybrn1.4.dat};
        \addplot [thin, gray] file {data/rt_surface_t-1.5_phi1.0_ybrn1.4.dat};
    \end{axis}
    \begin{scope}[xshift=4cm, yshift=0cm]
    \draw (0,0) -- (0,4cm);
    \draw[dashed] (0,0) -- (-3.78,3.78);
    \draw[dashed] (0,4) -- (-3.78,0.22);
    \draw[very thick] (-3.78,0)  -- (-3.78,4);
    \draw[gray, very thin] (0,0) -- (-3.78,0);
    \draw[gray, very thin] (0,4) -- (-3.78, 4);
    \draw[->] (0.5,1.5) -- node [above, rotate=-90] {$\tau_\bdry$}(0.5,2.5);
    \draw[->] (-4.28,1.5) -- node [above, rotate=90] {$\tau_\brane$}(-4.28,2.5);
    \end{scope}
    \end{tikzpicture}}
    \subcaptionbox{$\alpha = 1.0$\label{fig:rt_other_universes}}[0.45\textwidth]{
    \begin{tikzpicture}
    \begin{axis}[x=1cm, y=1cm, xticklabels=\empty, yticklabels=\empty, xmin=-2, xmax=2, ymin=-3, ymax=2, hide axis]
        \addplot [thick, red] file {data/rt_surface_t-1.8_phi1.0_ybrn1.4.dat};
        \addplot [thin, gray] file {data/rt_surface_t-2.1_phi1.0_ybrn1.4.dat};
        \addplot [thin, gray] file {data/rt_surface_t-2.4_phi1.0_ybrn1.4.dat};
        \addplot [thin, gray] file {data/rt_surface_t-2.7_phi1.0_ybrn1.4.dat};
        \addplot [thick, red] file {data/rt_surface_t-3.0_phi1.0_ybrn1.4.dat};
    \end{axis}
    \begin{scope}[xshift=4cm, yshift=3cm]
    \draw (0,1) -- (0,-3);
    \draw[dashed] (0,-2) -- (-3,1);
    \draw[dashed] (-1,1) -- (-3.78,-1.78);
    \draw[very thick] (-3.78,-3)  -- (-3.78,1);
    \draw[gray, very thin] (0,-2) -- (-3.78,-2);
    \draw[->] (0.5,-0.5) -- node [above, rotate=-90] {$\tau_\bdry$}(0.5,0.5);
    \draw[->] (-4.28,-0.5) -- node [above, rotate=90] {$\tau_\brane$}(-4.28,0.5);
    \draw[dashed] (0,-2) -- (-1,-3);
    \draw[dashed] (-3,-3) -- (-3.78,-2.22);
    \end{scope}
    \end{tikzpicture}}
    \caption{A family of RT surfaces emanating from the brane at different times. For (a), (b), (c) the red RT surfaces start at $\tau_\brane = 0, \pm 1.2$ and $y_\brane = -1.4$. (a) For $\alpha = 0.1$, the RT surfaces get deformed compared to case of vanishing DGP coupling in \cref{fig:two_dimensions_vanilla}, such that their ends are now at $\tau_\bdry < \tau_\brane$. (b) This trend becomes more prominent at $\alpha = 0.5$. (c) At $\alpha = 1.0$ we see that some extremal surfaces leave the patch of interest and end on the asymptotic boundary of a ``previous universe.'' (d) For large enough $\alpha$ there are other extremal surfaces which end on the brane in the analytic continuation of the spacetime. The red surfaces start at $\tau_\brane = -1.8, -3.0$.}
    \label{fig:rt_surfaces_non-zero_charge}
\end{figure}

The connected extremal surface trajectory is of course still given by \cref{eq:rt_trajectory_2d}. In \cref{fig:rt_surfaces_non-zero_charge} we display a few examples of connected extremal surfaces at different times for positive $\alpha$. The case of negative $\alpha$ is easily obtained by realizing that the image is invariant under a simultaneous $\mathbb Z_2$ transformation $\tau \to -\tau$, $\alpha \to - \alpha$.
In \cref{fig:rt_phi1} we see that for $\alpha = 1.0$ the extremal surfaces anchored at (roughly) negative values of $\tau_\brane$ leave the coordinate patch under consideration. More generally, this behaviour appears at a sufficiently large value of $|\alpha|$. Since there are no true singularities in AdS${}_3$, this naturally raises the question whether there is any meaning to the extremal surfaces which do not end on the asymptotic boundary within $\tau_\bdry \in (-\frac \pi 2, \frac \pi 2)$, but fall into the singularity. The answer is that such extremal surfaces compute entropies of subregions on the asymptotic boundary of the ``previous universe'' (or the ``next'' universe for negative $\alpha$) in the analytically continued spacetime.

This opens up the possibility of having additional extremal surfaces end on the asymptotic boundary within $\tau_\bdry \in (-\frac \pi 2, \frac \pi 2)$, which connect to the brane in the analytically continued spacetime, $\tau_\brane < - \frac{\pi}{2}$. This possibility is indeed realized, as depicted in \cref{fig:rt_other_universes}. For a more detailed discussion, refer to \cref{app:universe_crossing}. In fact, given a boundary time $\tau_\bdry \in (-\frac \pi 2, \frac \pi 2)$, there can generally be up to three candidate extremal surfaces which end on the brane. One of these will end on the brane at $\tau_\brane \in (-\frac \pi 2, \frac \pi 2)$, while the other two will end on the brane at $\tau_\brane < -\frac \pi 2$ ($\tau_\brane > \frac \pi 2$) for $\alpha$ positive (negative).

\begin{figure}
    \centering
    \subcaptionbox{$y_\brane=-0.1$}[0.45\textwidth]{
    \begin{tikzpicture}
    \begin{axis}[xlabel=$\tau_\brane$,ylabel=$\tau_\bdry$,ylabel shift = -0.2cm,
    xmin=-1.57,xmax=1.57,ymin=-1.57,ymax=1.57, x=1.7cm, y=1.7cm, legend pos=north west, legend style={draw=none}, legend cell align={left}]
    \addplot [thick, teal] file {data/phi0_y0.1.dat};
    \addplot [thick, olive] file {data/phi0.25_y0.1.dat};
    \addplot [thick, orange] file {data/phi0.5_y0.1.dat};
    \addplot [thick, red] file {data/phi0.75_y0.1.dat};
    \addplot [thick, purple] file {data/phi1.0_y0.1.dat};
    \addlegendentry{$\alpha=0.0$};
    \addlegendentry{$\alpha=0.25$};
    \addlegendentry{$\alpha=0.5$};
    \addlegendentry{$\alpha=0.75$};
    \addlegendentry{$\alpha=1.0$};
    \end{axis}
    \end{tikzpicture}
    }
    \subcaptionbox{$y_\brane=-1.4$\label{fig:tau_brane_vs_tau_boundary_y-1.4}}[0.45\textwidth]{
    \begin{tikzpicture}
    \begin{axis}[xlabel=$\tau_\brane$,ylabel=$\tau_\bdry$,ylabel shift = -0.2cm,
    xmin=-1.57,xmax=1.57,ymin=-1.57,ymax=1.57, x=1.7cm, y=1.7cm, legend pos=north west, legend style={draw=none}, legend cell align={left}]
    \addplot [thick, teal] file {data/phi0_y1.4.dat};
    \addplot [thick, olive] file {data/phi0.25_y1.4.dat};
    \addplot [thick, orange] file {data/phi0.5_y1.4.dat};
    \addplot [thick, red] file {data/phi0.75_y1.4.dat};
    \addplot [thick, purple] file {data/phi1.0_y1.4.dat};
    \addplot [thick, red, dashed] file {data/phi0.75_y1.4_extra.dat};
    \addplot [thick, purple, dashed] file {data/phi1.0_y1.4_extra.dat};
    \addlegendentry{$\alpha=0.0$};
    \addlegendentry{$\alpha=0.25$};
    \addlegendentry{$\alpha=0.5$};
    \addlegendentry{$\alpha=0.75$};
    \addlegendentry{$\alpha=1.0$};
    \end{axis}
    \end{tikzpicture}
    }
    \caption{The relation between the time $\tau_\brane$ at which a connected extremal surface connects to the brane and the time $\tau_\bdry$ at which it connects to the asymptotic boundary. (a) For $y_\brane = -0.1$, we see that even for $\alpha = 1.0$ the map between $\tau_\bdry$ and $\tau_\brane$ is one-to-one and thus all extremal surfaces connect the brane and boundary in the same universe. (b) For $y_\brane = -1.4$, a value of $\alpha \geq 0.75$ is sufficient for extremal surfaces to connect to other universes. The $\tau$ coordinate for the extremal surfaces of the dashed lines is $\tilde{\tau}_\brane = \tau_\brane - \pi$.}
    \label{fig:tau_brane_vs_tau_bdry}
\end{figure}
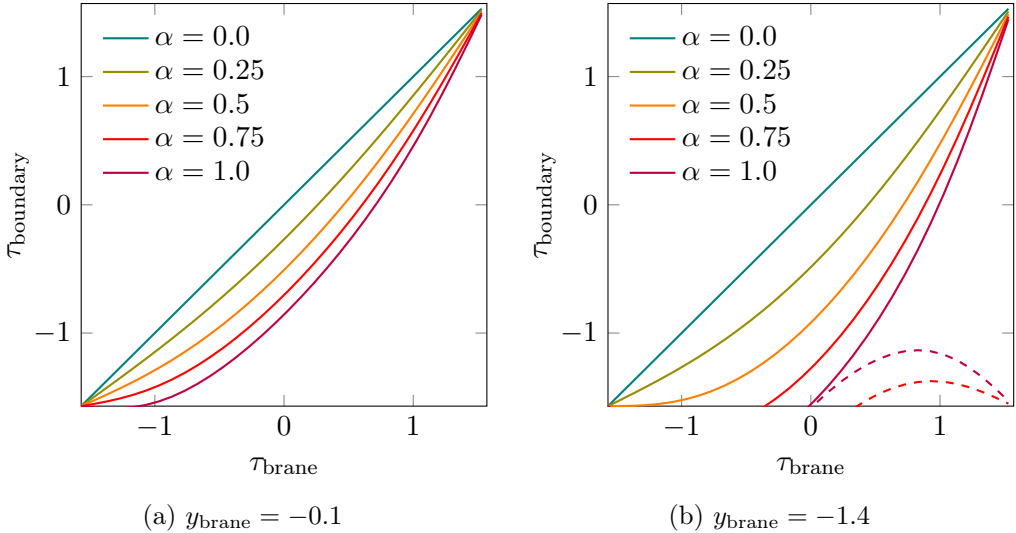

This becomes particularly clear once we examine the relation between the boundary and brane time. While in \cref{sec:subcr_vanishing_coupling} (with $\mathcal{Q}_E = 0$) the times at which a connected extremal surface intersected the asymptotic boundary and the brane were equal, i.e., $\tau_\brane = \tau_\bdry$, for non-zero $\mathcal Q_E$ we have from \cref{eq:rt_trajectory_2d} that
\begin{align}
    \label{eq:rt_trajectory_2d_alt}
    \tau_\bdry = \tau_\brane - \arcsin \left(\frac{\mathcal Q_E}{\sqrt{1 + \mathcal Q_E^2}} \sin y_\brane \right) + \arcsin\left(\frac{\mathcal Q_E}{\sqrt{1 + \mathcal Q_E^2}}\right).
\end{align}
Again, we stress that $\mathcal Q_E$ is a function of $\tau_\brane$, \cref{eq:QEalpha}. \Cref{fig:tau_brane_vs_tau_bdry} shows the relationship between $\tau_\bdry$ and $\tau_\brane$ for various values of $\alpha > 0$. The plots for negative $\alpha$ are obtained by reflection at the origin, i.e., $\tau_\brane \to - \tau_\brane$ and $\tau_\bdry \to - \tau_\bdry$. It is clear from the purple curve in \cref{fig:tau_brane_vs_tau_boundary_y-1.4} that, e.g., for $\alpha = 1.0$ we have three extremal surfaces which are connected to the brane for $\tau_\bdry \lesssim -1$. Two of these end on the brane approximately in the interval $\tau_\brane \in (- \pi, -\frac \pi 2)$.

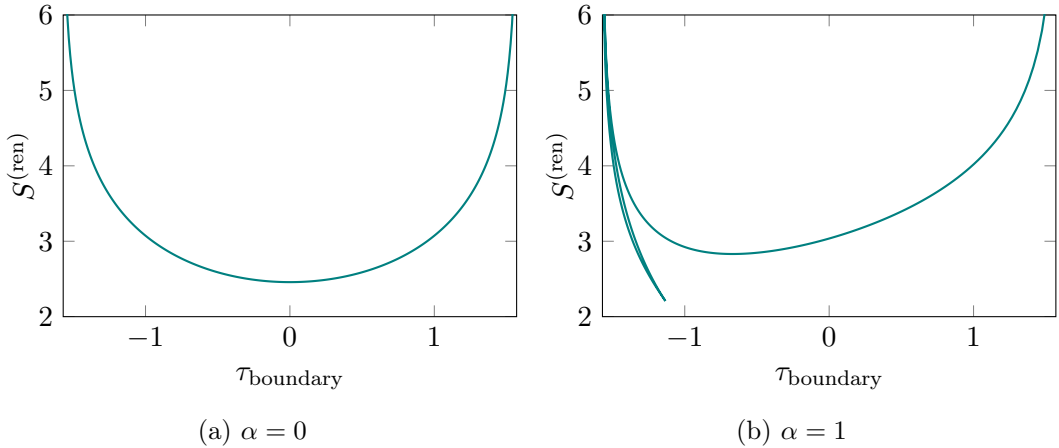
\begin{figure}
    \centering
    \subcaptionbox{$\alpha=0$}{
    \begin{tikzpicture}
    \begin{axis}[xlabel=$\tau_\bdry$,ylabel=${S}^\text{(ren)}$, ylabel shift = -0.2cm, xmin=-1.57, xmax=1.57, ymin=2, ymax=6, x=1.9cm, y=1cm]
    \addplot [thick, teal] file {data/rt_in_tau_phi0.0_y-1.4.dat};
    \end{axis}
    \end{tikzpicture}}
    \subcaptionbox{$\alpha=1$\label{fig:discontinuity_rt_surfaces_alpha_1}}{
    \begin{tikzpicture}
    \begin{axis}[xlabel=$\tau_\bdry$,ylabel=${S}^\text{(ren)}$, ylabel shift = -0.2cm, xmin=-1.57, xmax=1.57, ymin=2, ymax=6, x=1.9cm, y=1cm]
    \addplot [thick, teal] file {data/rt_in_tau_phi1.0_y-1.4.dat};
    \end{axis}
    \end{tikzpicture}}
    \caption{Renormalized entropy $ S^\text{(ren)}$ associated to extremal surfaces connecting to the brane. In both plots $y_\brane = -1.4$. (a) shows the situation for $\alpha = 0$. (b) For $\alpha=1$, we see that at early times there are three candidate surfaces. Of these, the surfaces which connect to the brane at much earlier times give the smallest entropy. Those surfaces are not available anymore after $\tau_\bdry \simeq -1.2$, and therefore the area of the smallest connected extremal surface exhibits a discontinuous jump.}
    \label{fig:discontinuity_rt_surfaces}
\end{figure}

An immediate question is which of the extremal surfaces has the minimum area, and hence yields the correct entanglement entropy. We will only answer the question here qualitatively and leave a quantitative analysis for the next subsection. \Cref{fig:discontinuity_rt_surfaces} shows the renormalized entropy $S^\text{(ren)}$, which equals $S^\text{(conn)}_{\text{vN}}$ renormalized by subtracting $S^\text{(th)}_{\text{vN}}$ and a constant which depends on various momentarily unimportant quantities, such as the constant contributions to the dilaton $\Phi_0$, $\varphi_0$, as well as the size of the boundary region $A$ under consideration.\footnote{Note that this also means that there is no contradiction between the fact that $S^\text{(ren)}$ is positive in, e.g., \cref{fig:discontinuity_rt_surfaces} and that the connected surface is the RT surface.} Recall that $S^\text{(th)}_{\text{vN}}$ is given by \cref{eq:exterior_entropy_2d} and $S^\text{(conn)}_{\text{vN}}$ is given by \cref{eq:s_dgp_2d}, with $\tau(y)$ related to the charge $\mathcal{Q}_E$ via \cref{eq:conserved_charge_2d,eq:rt_trajectory_2d}, implying that
\begin{equation}
    S^{\text{(ren)}} = \frac{L}{2G_N} \left[\alpha \sin(\tau_\brane) + \log\left(\frac{1}{\cos(\tau_\bdry)\sqrt{1+\mathcal{Q}_E^2}}\right) - \text{arcsinh}\left(\frac{\tan y_{\text{brane}}}{\sqrt{1+\mathcal{Q}_E^2}}\right)\right].
    \label{eq:sint_full_2d}
\end{equation}

\begin{figure}
    \centering
    \subcaptionbox{$y_\brane = -0.1$}{
    \begin{tikzpicture}
    \begin{axis}[xlabel=$t_\bdry$,ylabel=${S}^\text{(ren)}$, ylabel shift = -0.3cm, xmin=-6, xmax=6, ymin=-5, ymax=7, x=0.5cm, y=0.5cm, legend pos=south east, legend style={draw=none}, legend cell align={left}]
    \addplot [thick, teal] file {data/linear_growth_phi00_y-0.1.dat};
    \addplot [thick, olive] file {data/linear_growth_phi20_y-0.1.dat};
    \addplot [thick, orange] file {data/linear_growth_phi50_y-0.1.dat};
    \addplot [olive, thick, dotted] file {data/linear_growth_phi20_y-0.1_extra.dat};
    \addplot [orange, thick, dotted] file {data/linear_growth_phi50_y-0.1_extra.dat};
    \addlegendentry{$\alpha=0$};
    \addlegendentry{$\alpha=20$};
    \addlegendentry{$\alpha=50$};
    \end{axis}
    \end{tikzpicture}}
    \subcaptionbox{$y_\brane = -0.5$}{
    \begin{tikzpicture}
    \begin{axis}[xlabel=$t_\bdry$,ylabel=${S}^\text{(ren)}$, ylabel shift = -0.3cm, xmin=-6, xmax=6, ymin=-5, ymax=7, x=0.5cm, y=0.5cm, legend pos=south east, legend style={draw=none}, legend cell align={left}]
    \addplot [thick, teal] file {data/linear_growth_phi00_y-0.5.dat};
    \addplot [thick, olive] file {data/linear_growth_phi20_y-0.5.dat};
    \addplot [thick, orange] file {data/linear_growth_phi50_y-0.5.dat};
    \addplot [olive, thick, dotted] file {data/linear_growth_phi20_y-0.5_extra.dat};
    \addplot [orange, thick, dotted] file {data/linear_growth_phi50_y-0.5_extra.dat};
    \addlegendentry{$\alpha=0$};
    \addlegendentry{$\alpha=20$};
    \addlegendentry{$\alpha=50$};
    \end{axis}
    \end{tikzpicture}}
    \caption{Renormalized entropy $S^\text{(ren)}$ associated to extremal surfaces connecting to the brane as a function of the boundary time $t$. The entropies computed using extremal surfaces connecting to $\tau_\brane < - \frac \pi 2$ are displayed as dotted lines.}
    \label{fig:ve_violation_rt_surfaces}
\end{figure}
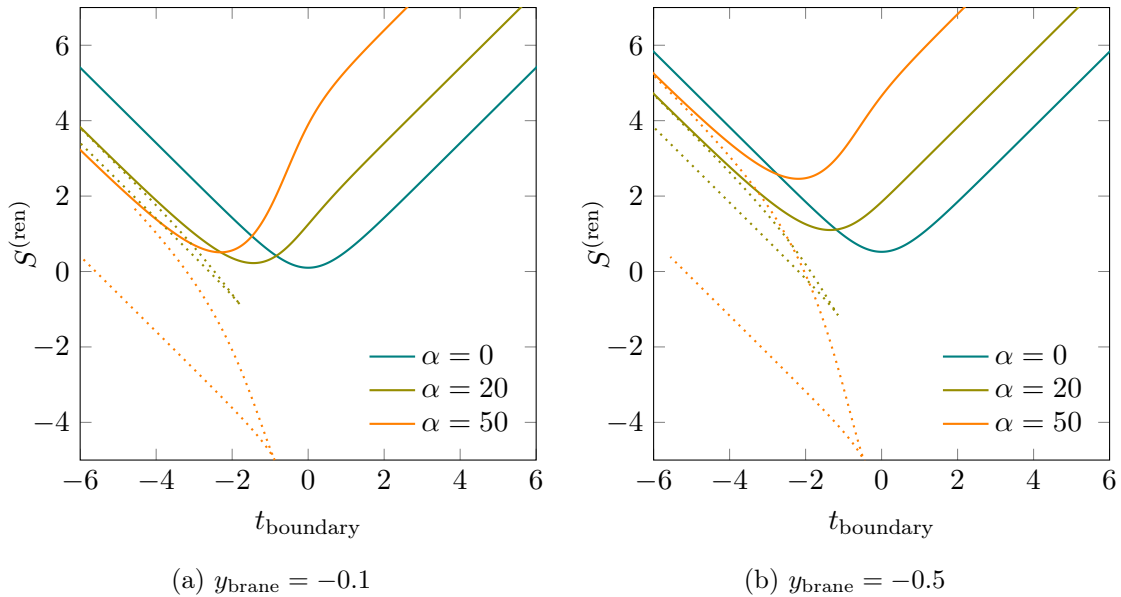

We see from \cref{fig:discontinuity_rt_surfaces_alpha_1,fig:ve_violation_rt_surfaces} that at early times there are three possible values for $S^\text{(ren)}$, which correspond to the areas of the three extremal surfaces which connect to the brane. The area associated with one of the extremal surfaces which connect to the brane in the previous universe is the minimum amongst all the connected extremal surfaces. It thus gives the correct entanglement entropy, unless $S^\text{(th)}_{\text{vN}}$ is even smaller. However, by choosing an arbitrarily large region $A$ we can always arrange for $S^\text{(th)}_{\text{vN}}$ to become as large as we like. Consequently, for sufficiently large regions, and sufficiently large $\alpha$, the holographic computation exhibits a discontinuity in the entanglement entropy. This is unphysical and the corresponding values of $\alpha$ need to be excluded.

Although extremal surfaces which connect to the previous (or next) universe exist, one might wonder if they could just be discarded, e.g., by requiring that the extremal surfaces must not leave the coordinate patch $\tau \in (-\frac \pi 2, \frac \pi 2)$. In \cref{fig:ve_violation_rt_surfaces} we have plotted $S^{(\text{ren})}$, now as a function of the Schwarzschild boundary time $t \in (-\infty, \infty)$, where $t$ is related to the boundary time $\tau_{\text{boundary}}$ via \begin{equation}\label{eq:t_to_tau}
t = \frac{L^2}{2r_+} \log\left(\frac{1+\sin \tau_{\text{boundary}}}{1-\sin \tau_{\text{boundary}}}\right).
\end{equation}
As in the previous plots, the situation for $\alpha < 0$ can be obtained by a reflection $t_\bdry \to - t_\bdry$. Disregarding extremal surfaces which connect to the previous universe, we can focus on the solid lines. The blue curves correspond to a vanishing dilaton. Apart from a change in the location of the minimum of $S^{(\text{ren})}$, we see that as we increase $\alpha$, the entropy develops a phase of very fast growth. This effect is more pronounced for small brane position $y_\brane$. We will now show that this fast growth of entanglement entropy is inconsistent from the boundary field theory perspective, such that either way we need to exclude a certain range of values of $\alpha$.

\subsection{Bounds on the coupling}
\label{sec:DGP_bounds_subcrit}
We have seen in the previous subsection that there are two effects which can potentially lead to an unphysical behaviour in the entanglement entropy for the CFT. These are
\begin{enumerate}
    \item A very rapid growth in entanglement, 
    \item A discontinuous jump in the entanglement entropy, due to extremal surfaces connecting to the previous/next universe.
\end{enumerate}
Both effects are at odds with the instantaneous bound on entanglement growth discussed in \cref{sec:limits_on_ve}, \cref{eq:ve_bound_2d}. If a holographic computation violates this bound, we can conclude that the bulk geometry cannot be dual to any physical state in the CFT, enabling us to put limits on the allowed range of values for the coupling $\alpha$.

We will start by considering the early time fast growth of entanglement entropy shown in the orange curves of \cref{fig:ve_violation_rt_surfaces}. In order to determine whether at any point in time we have a violation of the bound \cref{eq:ve_bound_2d}, it turns out to be sufficient to consider the late time behaviour of $v_E(t)$. If, for a particular choice of the JT coupling $\alpha$, $v_E$ approaches its asymptotic value of unity from above, it automatically implies that it must have exceeded unity by a finite amount during the course of its time evolution, violating \cref{eq:ve_bound_2d}. The corresponding JT coupling can thus be deemed unphysical.

For arbitrarily large $\alpha$ the RT surface at $\tau_\bdry \to \frac \pi 2$ will connect to $\tau_\brane \to \frac \pi 2$. The leading order terms in the asymptotic expansion of the entanglement velocity at $\tau_\bdry \to \frac \pi 2$, expressed in terms of the time at which the RT surface leaves the brane, $\tau_\brane$, is  
\begin{align}
    \label{eq:asymptotic_exp_2d}
    v_E(\tau) \sim 1 - \frac 1 2 (1 + \alpha - \alpha \sin y_\brane)(1 - \alpha - \alpha \sin y_\brane) \left(\frac \pi 2 - \tau_\brane\right)^2 + \ldots
\end{align}
The direction from which $v_E(\tau)$ asymptotically approaches unity is controlled by the sign of the second term in \cref{eq:asymptotic_exp_2d}. Given that we are considering the case $\alpha > 0$, the leading order correction to unity is negative if and only if $(1 - \alpha - \alpha \sin y_\brane) > 0$, or equivalently 
\begin{align}
    \label{eq:weak_bound_2d_1}
    \alpha < \frac 1 {1 + \sin y_\brane}.
\end{align}The asymptotic expansion at $\tau_\bdry \to -\frac \pi 2$ does not give a tighter bound when $\alpha > 0$.

For $\alpha < 0$, the bound is now given by the asymptotic expansion at $\tau_\bdry \to -\frac \pi 2$. It results in the same bound as in \cref{eq:weak_bound_2d_1}, now however for $- \alpha$. In summary, the bound on $\alpha$ by not allowing ultrafast (but finite) entanglement growth is 
\begin{align}
    \label{eq:weak_bound_2d}
    |\alpha| < \frac 1 {1 + \sin y_\brane}.
\end{align}
We are interested in the case where $-\frac{\pi}{2} < y_\brane \le 0$, and so the tightest bound is obtained for zero-tension branes, $y_\brane = 0$, where one has $|\alpha| < 1$. This bound, however, gets progressively weaker as the brane moves towards the left asymptotic boundary $y_\brane \to - \frac \pi 2$.

If we also consider RT surfaces which go through the singularity, a stronger bound can be obtained by requiring the absence of discontinuities in the entanglement entropy. Their absence is of course implied by the bounds on entanglement velocity, but should also be intuitively clear. In a CFT, entanglement propagates via local interactions, and since we evolve a translation-invariant state, the entanglement entropy cannot jump instantaneously by a finite amount at any point in time.

In order to better understand the conditions under which additional RT surfaces appear, notice that their presence implies that the map between $\tau_\brane$ and $\tau_\bdry$, \cref{eq:rt_trajectory_2d_alt}, is not one-to-one anymore. In the transition from a bijective to a surjective map, the function $\tau_\bdry(\tau_\brane)$ develops a saddle point at $\tau_\bdry = \tau_\brane = -\frac \pi 2$, c.f., \cref{fig:tau_brane_vs_tau_boundary_y-1.4}. For parameters above this saddle point value, the map will not be bijective anymore, and additional RT surfaces appear.
This condition can be used to find the physical values for the JT coupling by finding solutions to 
\begin{align}
    \partial_\tau \tau_\bdry(\tau_\brane) = 0
\end{align}
as a function of $y_\brane$ and $\alpha$. In fact, for positive coupling it is sufficient to find a solution to the above equation at $\tau_\brane = - \frac \pi 2$, since, as can be seen from \cref{fig:ve_violation_rt_surfaces}, the additional RT surfaces appear only at early times.
The resulting bound is
\begin{align}
    \alpha \leq  \frac 1 {1 - \sin y_\brane} \text{ for } \alpha \geq 0.
\end{align}

We can run the same argument for negative coupling, where we have to find a solution for the above equation at $\tau_\brane = \frac \pi 2$. This yields
\begin{align}
    \alpha \geq \frac 1 {-1 + \sin y_\brane} \text{ for }\alpha \leq 0.
\end{align}

Combining the two results, we can deduce the bound
\begin{align}
     |\alpha| \leq  \frac 1 {1 - \sin y_\brane}.
     \label{eq:alpha_bound}
\end{align}
For the regime of interest, i.e., $-\frac{\pi}{2} < y_\brane \leq 0$, this bound is always tighter than the bound \cref{eq:weak_bound_2d}. In order to argue that the parameter regime disallowed by \cref{eq:alpha_bound} needs to be excluded, we need to argue that there is in fact a discontinuous transition, i.e., these additional extremal surfaces are always smaller than the naive RT surface. In fact, as shown in \cref{app:proof}, we can make an even stronger statement: the first/last extremal surface to pass through the singularity is always the smallest one.

Of course, the question arises whether a configuration with the extremal surface passing through the singularity should be taken seriously. If we treat the AdS$_3$ case as a toy model for higher-dimensional cases, then perhaps we should ignore extremal surfaces which pass through the singularity altogether. A more compelling reason to exclude these extremal surfaces is that the entanglement wedge changes orientation in the ``previous universe.'' One the other hand, AdS${}_3$ does not have genuine curvature singularities, so nothing prevents us from continuing the solution past the apparent singularity and the orientation reversal is ``hidden'' in the singularity.

It would therefore be interesting to find an independent argument in favour of the stronger bound \cref{eq:alpha_bound}. We will leave this to future work. The only additional remark is that the situation discussed here, namely, that there are multiple extremal surfaces which connect the brane with the asymptotic boundary at a given boundary time, can also appear in higher dimensions. In that case, however, they appear far away from the singularity.



\section{Critical branes}
\label{sec:critical_branes}
As discussed in \cref{sec:brane_solutions}, there is another set of brane trajectories, now with $T_0 = T_\text{crit} = 1$, that cut off the left asymptotic boundary, c.f., \cref{fig:brane_trajectories}. We also discussed in \cref{sec:brane_solutions} how one can arrive at the critical case from the subcritical case by sending $T_0 \to 1$ while also shifting the Schwarzschild time by an appropriate amount simultaneously. Here, we will focus on a particular critical solution with the brane profile given by
\begin{align}
    \label{eq:new_sol_2d}
    \cos y_\brane + \sin y_\brane  = \mp \sin \tau_\brane,
\end{align}
which follows from \cref{eq:critical_family} after setting $t_0 = 0$, and is related to other critical solutions by time translations. This is of course justified since, as we have seen in the previous section, bounds on entanglement velocity bound the derivative of entanglement entropy and are insensitive to time translations. The upper sign describes a brane which comes out of the past singularity at $\tau = - \frac \pi 2$ and approaches the asymptotic boundary at $\tau = \frac \pi 2$. The lower sign is the time-reflected trajectory.

The brane trajectory $y(\tau)$ can easily be obtained from \cref{eq:new_sol_2d} and reads
\begin{align}
    \label{eq:new_sol_y_vs_tau}
    y_\brane = \text{arctan}\left(\frac { \sin \tau \pm \sqrt{2 - \sin^2 \tau} }{\sin \tau \mp \sqrt{2 - \sin^2 \tau} } \right).
\end{align}
We can now study extremal surfaces ending on this brane, first in the absence, and then in the presence of JT gravity on the brane.

\subsection{Branes with critical tension}
The correct solution for bulk extremal surfaces is still given by \cref{eq:rt_trajectory_2d}, and so we are only left with fixing the charge $\mathcal Q_E$. Since extremal surfaces which connect to the brane do not end at constant $y_\brane$ anymore, extremizing the area functional \cref{eq:area_functional_2d} gives a boundary term from the variation of the integrand as well as the variation of the limit of integration; see \cref{app:variation} for details. To make the boundary terms cancel we require that
\begin{align}
    \dot y_\brane(\tau) = \tau'_\rt(y) \Big|_\text{brane},
\end{align}
where the dot denotes a derivative with respect to $\tau$ and the prime is a derivative with respect to $y$. This is merely the statement that at the  point of intersection between the brane and the extremal surface, their respective tangent vectors need to be orthogonal in the Lorentzian sense. 

Given a point $(\tau,y)$ on the brane, the extremal surface emanating from this location must thus have a charge given by
\begin{align}
    \label{eq:charge_t_1}
    \mathcal Q_E = \frac {\dot y_\brane(\tau)}{\cos(y) \sqrt{1 - \dot y_\brane(\tau)^2}} = \mp \sqrt{- 2 \tan y}.
\end{align}
Here, we have made use of \cref{eq:new_sol_2d} in the second step. For the critical brane trajectory \cref{eq:new_sol_2d}, the proper time on the brane is related to the $y$-coordinate of the brane as\footnote{The value for $\mathcal Q_E$ takes a very simple form when expressed in terms of the proper time on the brane, $\mathcal Q_E = - \lambda$.}
\begin{align}
    \lambda = \pm \sqrt{- 2 \tan y}.
\end{align}
For simplicity, we will now focus only on the upper sign, i.e., a brane which exits the past singularity and approaches the left asymptotic boundary in the far future.

\begin{figure}
    \centering
    \begin{tikzpicture}
    \begin{axis}[x=1cm, y=1cm, xticklabels=\empty, yticklabels=\empty, xmin=-2, xmax=2, ymin=-2, ymax=2, hide axis]
        \addplot [thin, gray] file {data/rt_surface_T1_y0.dat};
        \addplot [thin, gray] file {data/rt_surface_T1_y-0.05.dat};
        \addplot [thin, gray] file {data/rt_surface_T1_y-0.2.dat};
        \addplot [thick, red] file {data/rt_surface_T1_y-0.4.dat};
        \addplot [thin, gray] file {data/rt_surface_T1_y-0.75.dat};
        \addplot [thin, gray] file {data/rt_surface_T1_y-1.1.dat};
        \addplot [thin, gray] file {data/rt_surface_T1_y-1.4.dat};
        \addplot [thick, red] file {data/rt_surface_T1_y-1.57.dat};
        \addplot [very thick, black] file {data/brane_T1.dat};
    \end{axis}
    \begin{scope}[xshift=4cm, yshift=0cm]
    \draw (0,0) -- (0,4cm);
    \draw[dashed] (0,0) -- (-4,4);
    \draw[dashed] (0,4) -- (-2.6,1.4);
    \draw[gray, very thin] (0,0) -- (-2,0);
    \draw[gray, very thin] (0,4) -- (-4, 4);
    \draw[->] (0.5,1.5) -- node [above, rotate=-90] {$\tau_\bdry$}(0.5,2.5);
    \draw[->] (-4.28,1.5) -- node [above, rotate=90] {$\tau_\brane$}(-4.28,2.5);
    \end{scope}
    \end{tikzpicture}
    \caption{Extremal surfaces which end on the brane in the case of $T_0 = T_\text{crit}$. As can be seen, all such surfaces end at time $\tau_\bdry = - \frac \pi 2$, which does not correspond to a finite value for the Schwarzschild time $t$.}
    \label{fig:rt_surfaces_t_1}
\end{figure}
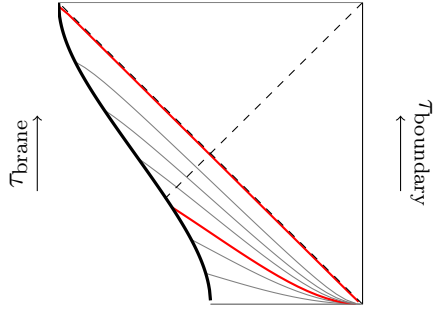

Using \cref{eq:rt_trajectory_2d_alt} together with \cref{eq:charge_t_1}, one can compute an explicit form for the bulk trajectory of extremal surfaces which end on the brane at $(y_\brane, \tau(y_\brane))$,
\begin{align}
    \tau_\rt(y) = \arcsin\left( \frac{\sqrt{1-2 \tan y_\brane \cos^2 y}-2 \tan y_\brane \sin y}{2 \tan y_\brane-1} \right).
\end{align}
In fact, as can be easily seen from this expression by setting $y = \frac \pi 2$, all extremal surfaces which leave the brane end at $\tau_\bdry = - \frac \pi 2$. This is shown in \cref{fig:rt_surfaces_t_1}. Consequently, there are no extremal surfaces which connect to the asymptotic boundary at a finite value for the Schwarzschild time. This makes it impossible for us to draw conclusions about the CFT state dual to the bulk geometry with a critical brane using bounds on entanglement velocity. Nonetheless, the introduction of JT gravity on the brane changes the behaviour of extremal surfaces, as we discuss in the next subsection. By introducing an appropriately chosen JT coupling, one can arrive at the situation where the extremal surfaces emanate from the brane and reach the asymptotic boundary at a finite value for the Schwarzschild time. Thus, we may still be able to place bounds on the coupling.

\subsection{Introducing JT gravity on the brane}
To analyse the effects of additional gravitational dynamics on the brane, we will focus on the simplest model of dilaton gravity on a Ricci-flat background, \cref{eq:JT_action} with $\Lambda^\brane = 0$. In addition to having this term in the brane action for the critical case, \cref{eq:braneaction} with $T_0 = 1$, we also allow for the possibility of an additional local coupling on the brane,
\begin{align}\label{eq:brane_counter}
    I_{\Delta T} = - \frac{1}{8\pi G^\brane_N}\int d^2x \sqrt{-h} \, \Delta T.
\end{align}
The equations of motion of the full system are given by
\begin{align}\label{eq:critical_eoms}
    R &= 0, &&
    \nabla_i \nabla_j \varphi = \frac{G_N^\brane}{G_N} \left(K_{ij} - \frac{1 + \Delta T}{L} h_{ij} \right),
\end{align}
where we have used that $T_0 = 1$. \Cref{eq:brane_eom} now implies $K_{ij} = \frac{1}{L} h_{ij}$, such that the right hand side of the second equation in \cref{eq:critical_eoms} simplifies to $- \frac{G_N^\brane}{G_N}  \frac{\Delta T}{L} h_{ij} $. Moreover, we once again require that $\varphi$ is invariant under translations in the transverse direction $x$. The general solution for the dilaton is then given by
\begin{align}
    \label{eq:dilaton_crit}
    \varphi(\lambda) = \frac{G_N^\brane}{2G_N}  \frac{\Delta T}{L}  \lambda^2 + c,
\end{align}
where $c$ is an arbitrary constant and $\lambda$ is the proper time on the brane.
As we can see, the dilaton profile is controlled by the parameter $\Delta T$ which plays a role similar to $\varphi_1$ in the previous section. We therefore define a new coupling
\begin{align}
    \alpha_\text{cr} \equiv \frac{G_N^\brane}{G_N}  \frac{\Delta T}{L},
\end{align}
which we attempt to constrain in the following discussion.

Requiring that the modified entropy functional, \cref{eq:rt_formula_w_dgp}, is extremized fixes the charge $\mathcal Q_E$ of an RT surface ending on the brane to be
\begin{align}
    \label{eq:charge_T_one}
    \mathcal Q_E = \left(\frac{\dot y}{\cos y \sqrt{1 - { \tau'_\rt} ^2}} - \partial_\tau \varphi(\tau) \right)\Bigg|_\brane.
\end{align}
Through the second term, the value of $\mathcal Q_E$ at any given time depends on the dilaton profile. Using the relation between $\tau$ and $\lambda$, one can easily compute that
\begin{align}
    \partial_\tau \varphi(\tau) \Big|_\brane = \alpha_\text{cr} \sqrt{\frac{- 2 \tan y}{1 - \sin (2 y)}}\Bigg|_\brane,
\end{align}
where we have made use of the brane trajectory \cref{eq:new_sol_2d}. In totality, \cref{eq:charge_T_one} takes the form
\begin{align}
    \label{eq:Qrt_T_1}
    \mathcal Q_E = -\sqrt{-2 \tan y} \left(\sqrt{1-2 \alpha_\text{cr} ^2 \tan y}+\alpha_\text{cr}   \sec y \sqrt{1-\sin (2 y)}\right)  \Big|_\brane.
\end{align}

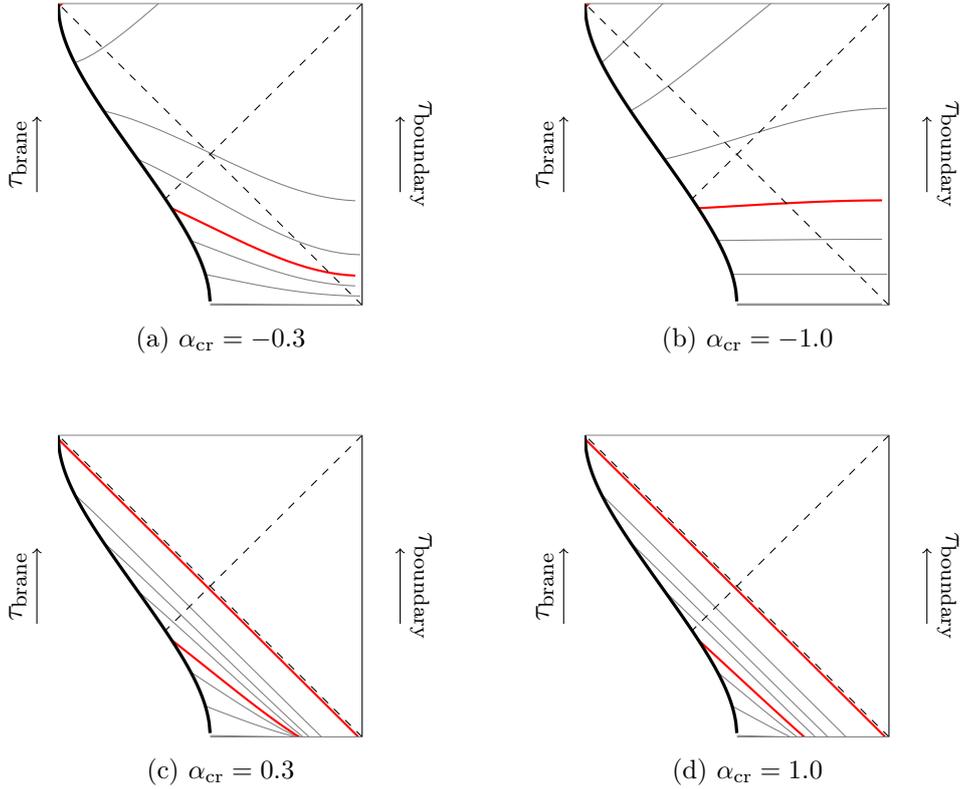
\begin{figure}
    \centering
    \subcaptionbox{\label{fig:rt_phi1_t1}$\alpha_\text{cr} = -0.3$}[0.45\textwidth]{
    \begin{tikzpicture}
    \begin{axis}[x=1cm, y=1cm, xticklabels=\empty, yticklabels=\empty, xmin=-2, xmax=2, ymin=-2, ymax=2, hide axis]
        \addplot [thin, gray] file {data/rt_surface_T1_y-0.0001_alpha_-0.3.dat};
        \addplot [thin, gray] file {data/rt_surface_T1_y-0.05_alpha_-0.3.dat};
        \addplot [thin, gray] file {data/rt_surface_T1_y-0.2_alpha_-0.3.dat};
        \addplot [thick, red] file {data/rt_surface_T1_y-0.4_alpha_-0.3.dat};
        \addplot [thin, gray] file {data/rt_surface_T1_y-0.75_alpha_-0.3.dat};
        \addplot [thin, gray] file {data/rt_surface_T1_y-1.1_alpha_-0.3.dat};
        \addplot [thin, gray] file {data/rt_surface_T1_y-1.4_alpha_-0.3.dat};
        \addplot [thick, red] file {data/rt_surface_T1_y-1.57_alpha_-0.3.dat};
        \addplot [very thick, black] file {data/brane_T1.dat};
    \end{axis}
    \begin{scope}[xshift=4cm, yshift=0cm]
    \draw (0,0) -- (0,4cm);
    \draw[dashed] (0,0) -- (-4,4);
    \draw[dashed] (0,4) -- (-2.6,1.4);
    \draw[gray, very thin] (0,0) -- (-2,0);
    \draw[gray, very thin] (0,4) -- (-4, 4);
    \draw[->] (0.5,1.5) -- node [above, rotate=-90] {$\tau_\bdry$}(0.5,2.5);
    \draw[->] (-4.28,1.5) -- node [above, rotate=90] {$\tau_\brane$}(-4.28,2.5);
    \end{scope}
    \end{tikzpicture}}
    \subcaptionbox{\label{fig:rt_other_universes_t1}$\alpha_\text{cr} = -1.0$}[0.45\textwidth]{
    \begin{tikzpicture}
    \begin{axis}[x=1cm, y=1cm, xticklabels=\empty, yticklabels=\empty, xmin=-2, xmax=2, ymin=-2, ymax=2, hide axis]
        \addplot [thin, gray] file {data/rt_surface_T1_y-0.0001_alpha_-1.0.dat};
        \addplot [thin, gray] file {data/rt_surface_T1_y-0.05_alpha_-1.0.dat};
        \addplot [thin, gray] file {data/rt_surface_T1_y-0.2_alpha_-1.0.dat};
        \addplot [thick, red] file {data/rt_surface_T1_y-0.4_alpha_-1.0.dat};
        \addplot [thin, gray] file {data/rt_surface_T1_y-0.75_alpha_-1.0.dat};
        \addplot [thin, gray] file {data/rt_surface_T1_y-1.1_alpha_-1.0.dat};
        \addplot [thin, gray] file {data/rt_surface_T1_y-1.4_alpha_-1.0.dat};
        \addplot [thick, red] file {data/rt_surface_T1_y-1.57_alpha_-1.0.dat};
        \addplot [very thick, black] file {data/brane_T1.dat};
    \end{axis}
    \begin{scope}[xshift=4cm, yshift=0cm]
    \draw (0,0) -- (0,4cm);
    \draw[dashed] (0,0) -- (-4,4);
    \draw[dashed] (0,4) -- (-2.6,1.4);
    \draw[gray, very thin] (0,0) -- (-2,0);
    \draw[gray, very thin] (0,4) -- (-4, 4);
    \draw[->] (0.5,1.5) -- node [above, rotate=-90] {$\tau_\bdry$}(0.5,2.5);
    \draw[->] (-4.28,1.5) -- node [above, rotate=90] {$\tau_\brane$}(-4.28,2.5);
    \end{scope}
    \end{tikzpicture}}\vspace{10mm}
        \subcaptionbox{$\alpha_\text{cr} = 0.3$}[0.45\textwidth]{
    \begin{tikzpicture}
    \begin{axis}[x=1cm, y=1cm, xticklabels=\empty, yticklabels=\empty, xmin=-2, xmax=2, ymin=-2, ymax=2, hide axis]
        \addplot [thin, gray] file {data/rt_surface_T1_y-0.0001_alpha_0.3.dat};
        \addplot [thin, gray] file {data/rt_surface_T1_y-0.05_alpha_0.3.dat};
        \addplot [thin, gray] file {data/rt_surface_T1_y-0.2_alpha_0.3.dat};
        \addplot [thick, red] file {data/rt_surface_T1_y-0.4_alpha_0.3.dat};
        \addplot [thin, gray] file {data/rt_surface_T1_y-0.75_alpha_0.3.dat};
        \addplot [thin, gray] file {data/rt_surface_T1_y-1.1_alpha_0.3.dat};
        \addplot [thin, gray] file {data/rt_surface_T1_y-1.4_alpha_0.3.dat};
        \addplot [thick, red] file {data/rt_surface_T1_y-1.57_alpha_0.3.dat};
        \addplot [very thick, black] file {data/brane_T1.dat};
    \end{axis}
    \begin{scope}[xshift=4cm, yshift=0cm]
    \draw (0,0) -- (0,4cm);
    \draw[dashed] (0,0) -- (-4,4);
    \draw[dashed] (0,4) -- (-2.6,1.4);
    \draw[gray, very thin] (0,0) -- (-2,0);
    \draw[gray, very thin] (0,4) -- (-4, 4);
    \draw[->] (0.5,1.5) -- node [above, rotate=-90] {$\tau_\bdry$}(0.5,2.5);
    \draw[->] (-4.28,1.5) -- node [above, rotate=90] {$\tau_\brane$}(-4.28,2.5);
    \end{scope}
    \end{tikzpicture}}
    \subcaptionbox{$\alpha_\text{cr} = 1.0$}[0.45\textwidth]{
    \begin{tikzpicture}
    \begin{axis}[x=1cm, y=1cm, xticklabels=\empty, yticklabels=\empty, xmin=-2, xmax=2, ymin=-2, ymax=2, hide axis]
        \addplot [thin, gray] file {data/rt_surface_T1_y-0.0001_alpha_1.0.dat};
        \addplot [thin, gray] file {data/rt_surface_T1_y-0.05_alpha_1.0.dat};
        \addplot [thin, gray] file {data/rt_surface_T1_y-0.2_alpha_1.0.dat};
        \addplot [thick, red] file {data/rt_surface_T1_y-0.4_alpha_1.0.dat};
        \addplot [thin, gray] file {data/rt_surface_T1_y-0.75_alpha_1.0.dat};
        \addplot [thin, gray] file {data/rt_surface_T1_y-1.1_alpha_1.0.dat};
        \addplot [thin, gray] file {data/rt_surface_T1_y-1.4_alpha_1.0.dat};
        \addplot [thick, red] file {data/rt_surface_T1_y-1.57_alpha_1.0.dat};
        \addplot [very thick, black] file {data/brane_T1.dat};
    \end{axis}
    \begin{scope}[xshift=4cm, yshift=0cm]
    \draw (0,0) -- (0,4cm);
    \draw[dashed] (0,0) -- (-4,4);
    \draw[dashed] (0,4) -- (-2.6,1.4);
    \draw[gray, very thin] (0,0) -- (-2,0);
    \draw[gray, very thin] (0,4) -- (-4, 4);
    \draw[->] (0.5,1.5) -- node [above, rotate=-90] {$\tau_\bdry$}(0.5,2.5);
    \draw[->] (-4.28,1.5) -- node [above, rotate=90] {$\tau_\brane$}(-4.28,2.5);
    \end{scope}
    \end{tikzpicture}} 
    \caption{A family of extremal surfaces emanating from the brane for the critical case. The red surfaces start at $y = -0.4$ and $y = -1.57$ on the brane. (a) and (b) indicate that for $\alpha_\text{cr} < 0$ extremal surfaces can end on the asymptotic boundary at $\tau_\bdry > - \frac{\pi}{2}$. (c) and (d) show that for $\alpha_\text{cr} > 0$ there are no extremal surfaces connecting the brane with the asymptotic boundary within the same coordinate patch.}
    \label{fig:rt_surfaces_T1_non-zero_charge}
\end{figure}

\subsection{Bounds on the coupling}
We can now use the result \cref{eq:Qrt_T_1} in the equations for the trajectory and the area of the connected surface to obtain the behaviour of an extremal surface which connects to the brane at $y = y_\brane$. The resulting extremal surfaces are shown in \cref{fig:rt_surfaces_T1_non-zero_charge}. 

It is obvious that there is a qualitative difference between $\alpha_\text{cr} < 0$ and $\alpha_\text{cr} > 0$. In the first case, shown in (a) and (b) of \cref{fig:rt_surfaces_T1_non-zero_charge}, there are extremal surfaces that end on the asymptotic boundary at $\tau_\bdry > -\frac{\pi}{2}$. On the contrary, as shown in (c) and (d), if $\alpha_\text{cr} > 0$, one does not find extremal surfaces which connect the brane to the asymptotic boundary. In other words, for $\alpha_\text{cr} > 0$, the von Neumann entropy is always given by the thermal value \cref{eq:exterior_entropy_2d}. We might thus at best hope for a lower bound on $\alpha_\text{cr}$.

\begin{figure}
    \centering
    \begin{tikzpicture}
    \begin{axis}[xlabel=$t_\bdry$,ylabel=${S}^{(\text{ren})}$, ylabel shift = -0.3cm, xmin=-6, xmax=6, ymin=-6, ymax=6, x=0.5cm, y=0.5cm, legend pos=south west, legend style={draw=none}, legend cell align={left}]
    \addplot [thick, teal] file {data/linear_growth_crit_alpha-0.2.dat};
    \addplot [thick, olive] file {data/linear_growth_crit_alpha-2.0.dat};
    \addplot [thick, orange] file {data/linear_growth_crit_alpha-6.0.dat};
    \addlegendentry{$\alpha=-0.2$};
    \addlegendentry{$\alpha=-2.0$};
    \addlegendentry{$\alpha=-6.0$};
    \end{axis}
    \end{tikzpicture}
    \caption{Renormalized entropy $S^{(\text{ren})}$ associated to extremal surfaces connecting to the brane as a function of the boundary time $t_\bdry$.}
    \label{fig:critical_schwarzschild}
\end{figure}
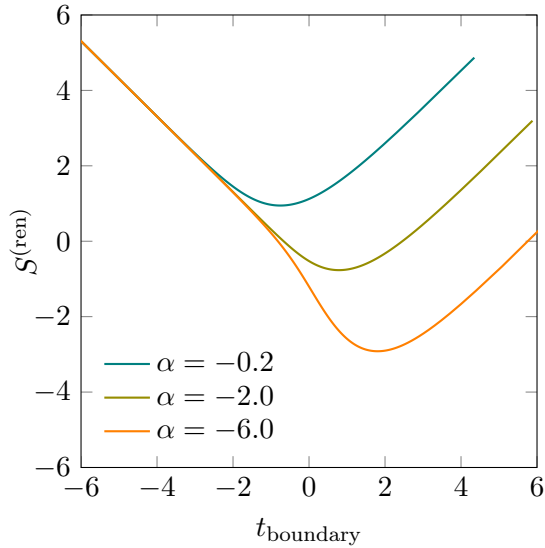

A comprehensive understanding can be obtained by plotting the entanglement entropy against the Schwarzschild time $t$, \cref{fig:critical_schwarzschild}. The plot shows that similar to the subcritical case, for certain values of $\alpha_\text{cr} < 0$ the entropy growth becomes very fast, and can violate the bound \cref{eq:ve_bound_2d}. The critical value for $\alpha_\text{cr}$ below which this happens can once again be obtained by using the asymptotic expansion of the entanglement velocity at early times,
\begin{align}
    |v_E(\tau)| \sim 1 + \frac{\alpha_\text{cr}}{2} (2 + \alpha_\text{cr}) \left(\tau_\brane + \frac \pi 2\right)^2 + \dots.
\end{align}
We thus conclude that states with physically acceptable entanglement velocity \cref{eq:ve_bound_2d} must have
\begin{align}
\label{eq:alpha_cr_bound_weak}
\alpha_\text{cr} \geq -2,
\end{align} 
while $\alpha_\text{cr}$ is unbounded from above, since in that case no extremal surfaces exist which connect the brane to the asymptotic boundary within the same coordinate patch.

In order to try to make statements about the states with $\alpha_\text{cr} > 0$, we can again consider as to what happens if we allow for extremal surfaces to pass through the singularity and end on the brane in the previous patch.
As can be seen from \cref{fig:rt_surfaces_T1_non-zero_charge}, we indeed have the possibility of extremal surfaces passing through the singularity. Similar to the case of subcritical branes, we can analytically continue our solution and consider the case where extremal surfaces can end on the resulting geometry behind the singularity i.e.\ in the previous universe. For the present case, we can only extend our solution to $\tau < -\frac \pi 2$, since the dilaton diverges towards the future. The continued solution is symmetric under $\tau \to \pi - \tau$. It turns out that unlike the subcritical case, extremal surfaces now cross the singularity for arbitrarily small $\alpha_\text{cr} > 0$ and end at $\tau_\bdry > - \frac{\pi}{2}$. Since the new extremal surfaces are the only extremal surfaces in the connected phase, they will dominate for sufficiently large boundary subregions. There is a critical time after which such surfaces cease to exist. Consequently, there will be a discontinuous jump in the entanglement entropy. Thus, if this approach of analytically continuing the geometry is to be trusted, $\alpha_\text{cr} > 0$ should be ruled out completely, and we arrive at the bound
\begin{align}
    \label{eq:alpha_cr_bound_strong}
    0 \geq \alpha_\text{cr} \geq -2.
\end{align}
Interestingly, repeating the analysis with the time reflected solution, we obtain the same bounds. This is however expected, since we arrive at the time-reflected solution by mapping $\tau \to - \tau$, but leaving $\alpha_\text{cr}$ invariant.

\subsection{Comments on bounds in the critical case}
Recall that in the subcritical case, the bound that followed by disallowing very fast entanglement growth, \cref{eq:weak_bound_2d}, allowed for arbitrarily large values of $|\alpha|$ as the brane approached the left asymptotic boundary. It might therefore be surprising that a strict bound is obtained in the critical case, which, as discussed earlier, can be related to the subcritical case in an appropriate limit. We now  explore the relation between the two cases in detail, as well as demonstrate how the bounds \cref{eq:alpha_cr_bound_weak,eq:alpha_cr_bound_strong} can be derived from the bounds of \cref{sec:subcritical_branes}.

We have already discussed in \cref{sec:brane_solutions} how the critical case can be obtained from the subcritical case by taking the brane towards the left asymptotic boundary, while at the same time performing a time translation such that \cref{eq:fixed} is held fixed. One can show via a lengthy computation that this transformation also relates the proper time in both cases, 
\begin{align}\label{eq:lam_subc_c}
    \lambda_\text{sub-cr} = - \frac{1}{\cos y_\brane} \frac \pi 2 + \lambda_\text{cr} + \mathcal O\left(\cos y_\brane\right).
\end{align}
We will refrain from giving a detailed derivation of this result. This is because it is much easier to arrive at it by realizing that \cref{eq:lam_subc_c} exactly maps the induced metric for the subcritical case, \cref{eq:induced_metric_T_less_1}, to the one for the critical case, \cref{eq:induced_metric_T_eq_1}, as we take $y_\brane \to -\frac \pi 2$.

Using this relation one can compare the dilaton solutions \cref{eq:dilaton_def} and \cref{eq:dilaton_crit} with one-another to arrive at\footnote{Note that in order to compare to the computation of the previous subsection, we have $\alpha_\text{cr}<0$ and thus also $\alpha < 0$.}
\begin{align}
    \label{eq:relation_alpha}
    \alpha_\text{cr} = 2 \, (1 + \sin y_\brane) \, \alpha.
\end{align}
This tells us that if we are interested in obtaining a critical brane with a non-trivial dilaton profile, we need to let $\alpha$ diverge as we take the subcritical brane to its critical limit. 
If we only consider the weak bound on $\alpha$, \cref{eq:weak_bound_2d}, we find that this is indeed possible, provided that the right hand side of \cref{eq:relation_alpha} is between $-2$ and $2$. On the other hand, if one considers the stronger bound in the subcritical case that follows from analytic continuation, \cref{eq:alpha_bound}, which requires that $\alpha$ is finite in the limit $y_\brane \to - \frac \pi 2$, then one can only have $\alpha_\text{cr} = 0$.

This limiting procedure in fact suggests a stronger bound than the one obtained from the RT surfaces alone. Given that in fact $|\alpha|$ was bounded, we can conclude that for solutions obtained by the limit explained above, 
\begin{align}
    | \alpha_\text{cr}| \leq 2,
\end{align}
from ultrafast entanglement growth, or 
\begin{align}
    \alpha_\text{cr} = 0,
\end{align}
from RT surfaces that pass through the singularity.


\section{Brane tomography}
\label{sec:tomography}
We have seen in the previous sections that adding JT gravity to the brane modifies the behaviour of the entanglement entropy for the dual CFT state as a function of time. We are now interested in a slightly different problem. Given a large interval on the boundary, which from the bulk perspective means that the connected extremal surfaces are the RT surfaces up to arbitrarily late times, what are the signatures of the presence of JT gravity on the brane, that can be extracted from the behaviour of the entanglement entropy as a function of time? In other words, we would like to understand, from a purely boundary perspective, the imprints of any intrinsic gravitational dynamics on the brane, and how these can by computed from the asymptotic behaviour of the holographic entanglement entropy. For the present case, this would correspond to reading off the value of the JT coupling $\alpha$ as well as the constant term in the solution for the dilaton, by looking at the asymptotic rate of entanglement growth. Also, in addition to these, we will be able to extract the location of the brane within the bulk geometry, $y_\brane$, along with the parameter $t_0$ that appears in the brane trajectory, \cref{eq:t_less_1_general_schwsch,eq:brane_traj_crit}, as we illustrate below. We call this procedure \emph{brane tomography}. 

\subsection{Subcritical brane tomography}
We first perform tomography on CFT states dual to bulk geometries with subcritical branes, discussed in detail in \cref{sec:DGP2D}. Given that we consider a very large interval on the boundary, the entanglement entropy is given by the connected extremal surface, \cref{eq:s_dgp_2d}. Making use of \cref{eq:def_cutoff_y,eq:rt_trajectory_2d,eq:dilaton_def,eq:def_alpha,eq:QEalpha}, it can be cast into the explicit form\footnote{For notational convenience we work in the units where $L/2G_N = 1$.}
\begin{equation}
\begin{split}
    S_{\text{vN}} = \,\,&\frac{\Phi_0 + \varphi_0}{2 G_N^{\brane}} +\alpha \sin \tau_{\brane} + \log\left(\frac{\beta}{\pi L \cos\tau_{\bdry}} \frac{1}{\sqrt{1+\alpha^2 \cos^2\tau_{\brane}}}\right)\\
    &- \text{arcsinh}\left(\frac{\tan y_{\brane}}{\sqrt{1+\alpha^2 \cos^2\tau_{\brane}}}\right) - \log \epsilon + \mathcal{O}(\epsilon^2),
    \label{eq:sint_full_2d_2}
    \end{split}
\end{equation}
with the regulator $\epsilon \rightarrow 0$.
As alluded to above, we will be interested in extracting the JT coupling $\alpha$ along with the brane parameters $(y_{\text{brane}}, t_0)$ by studying the asymptotic behaviour of the entanglement entropy as a function of the boundary time $t$. To do so, we look at the entanglement velocity $v_E$, defined in \cref{eq:def_ve}. As a first step in performing tomography, one can determine the sign of $\alpha$ by looking at the behaviour of $v_E$ around its minimum. For $\alpha>0$, the absolute value of the time rate of change of $v_E$ is larger after the minimum, compared to before reaching the minimum, as is evident from \cref{fig:ve_violation_rt_surfaces}. The opposite holds true for $\alpha < 0$. 

With the sign of $\alpha$ in hand, the next step is to examine the asymptotic behaviour of $v_E$. This requires some care, because whether one should look at the asymptotic behaviour for $t\rightarrow \infty$ or $t\rightarrow -\infty$ depends upon the sign of $\alpha$. This can be thought of as a consequence of \cref{eq:sint_full_2d_2}, which does not give the expression for the entanglement entropy directly as a function of the Schwarzschild time $t$, but rather in terms of $\tau_\brane$. For all $\alpha > 0$, the limit $\tau_\brane \rightarrow \frac{\pi}{2}$ corresponds to $t\rightarrow \infty$, as can be easily computed using \cref{eq:rt_trajectory_2d_alt,eq:t_to_tau}, and has also been depicted in \cref{fig:rt_surfaces_non-zero_charge}. The other asymptotic limit, $t \rightarrow -\infty$, is reached at a value of $\tau_\brane$ that for $\alpha > - (1 - \sin y_\brane)$ depends upon the explicit value of $\alpha$ and is thus much harder to extract.\footnote{For performing brane tomography, we restrict our attention to the coordinate patch $\tau_\bdry \in \left(-\frac{\pi}{2},\frac{\pi}{2}\right)$.} Thus, for $\alpha > 0$, one should look at the asymptotic expansion of $v_E$ as $t\rightarrow \infty$. On the other hand, when $\alpha < 0$, the correct extremal surfaces can be obtained by reflecting the extremal surfaces depicted in \cref{fig:rt_surfaces_non-zero_charge} about the horizontal $\tau_\brane = 0$ axis. Consequently, for such cases, the limit $\tau_\brane \rightarrow -\frac{\pi}{2}$ always leads to $t\rightarrow -\infty$, and therefore this is the appropriate point one should expand $v_E$ about when $\alpha < 0$.

In the following, we assume that $\alpha > 0$, unless otherwise stated. Then, using \cref{eq:sint_full_2d_2}, the asymptotic behaviour of $v_E$ as $t\rightarrow \infty$ takes the form
\begin{equation}\label{eq:asymptopia}
v_E = 1 + \mathcal{K}_1\,  e^{-\frac{4\pi t}{\beta}} + \mathcal{K}_2\, e^{-\frac{8\pi t}{\beta}} + \mathcal{K}_3 \, e^{-\frac{12\pi t}{\beta}} +\mathcal{O}\left(e^{-\frac{16\pi t}{\beta}}\right).
\end{equation}
To arrive at this expression, it is required to compute the time derivative of the entanglement entropy in the connected phase, \cref{eq:sint_full_2d_2}, with respect to the Schwarzschild time $t$. This can be done by using the chain rule for differentiation. Given that \cref{eq:sint_full_2d_2} expresses the entanglement entropy as a function of $\tau_\brane$ and $\tau_\bdry$, with $\tau_\bdry$ itself being a monotonically increasing function of $\tau_\brane$ (at least in the parameter regime of interest), \cref{eq:rt_trajectory_2d_alt}, it is straightforward to compute the derivative ${\partial S_{\text{vN}}}/{\partial \tau_\brane}$ as a function of $\tau_\brane$. Similarly, using \cref{eq:t_to_tau}, one can compute ${\partial t}/{\partial \tau_\brane}$ as a function of $\tau_\brane$. Combining the two yields ${\partial S_{\text{vN}}}/{\partial t}$, albeit as a function of $\tau_\brane$. Now, for all $\alpha > 0$, the late time limit $t\rightarrow \infty$ corresponds to $\tau_\brane \rightarrow \frac{\pi}{2}$, as discussed above. One can thus take the limit of $\tau_\brane \rightarrow \frac{\pi}{2}$ in the expression for $\partial S_{\text{vN}}/\partial t$, which gives the late time behaviour of $v_E$ as an expansion in powers of $\left(\tau_\brane - \frac{\pi}{2}\right)^2$. The final step then requires one to convert from $\tau_\brane$ to $t$ in the late time limit, which can be done using \cref{eq:t_to_tau}, yielding the asymptotic expansion \cref{eq:asymptopia}. 

The leading corrections to the asymptotic behaviour of $v_E$, parametrized by $\mathcal{K}_1, \mathcal{K}_2, \mathcal{K}_3 \ldots$ in \cref{eq:asymptopia}, are functions of the brane location $y_{\text{brane}}$ and the JT coupling $\alpha$. Using the notation $\xi \equiv 1 - \alpha \sin y_\brane$ for brevity, we find that
\begin{subequations}
\begin{align}
   &\mathcal{K}_1 = -\frac{2}{\left(\alpha+\xi\right)} \left[\xi - \alpha \right],\label{eq:tomo2d1}\\
    &\mathcal{K}_2 = \phantom{-}\frac{2}{\left(\alpha+\xi\right)^4} \left[\xi^4 - 2 \alpha\left(\alpha-2\right)\xi^2 - 4\alpha\xi +\alpha^3(\alpha-4)\right],\label{eq:tomo2d2}\\
    &\mathcal{K}_3 = - \frac{2}{\left(\alpha+\xi\right)^7} \Big[\xi^7 + \alpha \xi^6 - 3\alpha(\alpha-4) \xi^5 - 3 \alpha \left(\alpha^2 -4\alpha + 8\right)\xi^4 + 3\alpha(\alpha^3 - 8\alpha^2 +8) \xi^3 \nonumber \\
    &\hspace{28mm}+3\alpha(\alpha(\alpha-2)(\alpha^2-6\alpha + 4) - 4) \xi^2 - \alpha^2(\alpha^4 - 12 \alpha^3 +40\alpha -12)\xi \nonumber\\
    &\hspace{28mm}- \alpha^4 (\alpha-2)(\alpha^2 - 10\alpha+4)\Big], \label{eq:tomo2d3}
    \end{align}
\label{eq:tomo2d}
\end{subequations}
and so on. Thus, the $\mathcal{K}_i$ encapsulate the values of the brane parameters $(y_\text{brane},\alpha)$. Following the discussion in \cref{sec:DGP_bounds_subcrit}, demanding $\mathcal{K}_1 < 0$ to ensure that \cref{eq:ve_bound_2d} holds true provides a bound on the allowed values of the JT coupling $\alpha$, \cref{eq:weak_bound_2d_1}.

In the discussion above, we restricted our attention to the case $\alpha>0$. As discussed earlier, when $\alpha<0$, rather than looking at the asymptotic expansion for the entanglement velocity at very late times, one must instead consider the asymptotic expansion corresponding to very early times, i.e., $t \rightarrow -\infty$, where $v_E \rightarrow -1$. By repeating the analysis above, we once again get an asymptotic expansion for the entanglement velocity of the form given in \cref{eq:asymptopia}, albeit now with the replacements $v_E \rightarrow - v_E$, $t \rightarrow -t$ and $\alpha \rightarrow -\alpha$. This implies that our expressions for the functions $\mathcal{K}_i$ that were obtained assuming $\alpha > 0$ are also the correct coefficients for the asymptotic expansion around $t \to -\infty$ for $\alpha < 0$, provided we make the replacement $\alpha\rightarrow -\alpha$ in \cref{eq:tomo2d}.

It is tempting to think about the possibility of extracting the brane parameters $(y_{\text{brane}}, \alpha)$ using $\mathcal{K}_i$. However, there is one important subtlety here --- the quantities $\mathcal{K}_i$ are not invariant under boundary time translations. Consider two different coordinate systems on the boundary, such that their time coordinates are shifted relative to one another by an amount $\Delta t$. This would imply that the values for $\mathcal{K}_1$ between the two coordinate systems will differ by a factor of $e^{{-4\pi \Delta t}/{\beta}}$, the values for $\mathcal{K}_2$ will differ by a factor of $e^{{-8\pi \Delta t}/{\beta}}$, and so on.\footnote{In fact, by comparing the values for the same coefficient, for instance $\mathcal{K}_1$, between two different coordinate systems on the boundary, one can determine the relative time shift between them.} Therefore, to ensure that one can extract the same values for the brane parameters from the asymptotic behaviour of $v_E$, one should work with quantities that are agnostic to relative time shifts between different coordinate systems. We choose to work with the invariants
\begin{subequations}
\begin{align}
    \aleph_1 &\equiv \frac{\mathcal{K}_2 - \frac{1}{2} \mathcal{K}_1^2}{\mathcal{K}_1^2}\, , \label{eq:def_aleph_1}\\
    \aleph_2 &\equiv \frac{\mathcal{K}_3 - \frac{3}{2}\mathcal{K}_2 \mathcal{K}_1 + \frac{1}{2}\mathcal{K}_1^3}{\mathcal{K}_1^3}\, , \label{eq:def_aleph_2}
\end{align} \label{eq:the_alephs}
\end{subequations}
which, using \cref{eq:tomo2d}, have the explicit form
\begin{subequations}
\begin{align} 
    \aleph_1 &= \frac{2 \alpha }{(\alpha - \xi)^2 (\alpha + \xi)^2}\, \left[\xi^2-\xi-\alpha^2\right], \label{eq:aleph_1_xi} \\
    \aleph_2 &= \frac{\alpha}{(\alpha-\xi)^3 (\alpha+\xi)^4}\,\big[3\xi^4 -3(\alpha+2) \xi^3 -3(3\alpha^2 -2\alpha-1)\xi^2 \nonumber\\
    &\hspace{34mm}+ \alpha(3\alpha^2+10\alpha-3)\xi + 2\alpha^3(3\alpha-1)\big]. \label{eq:aleph_2_xi}
\end{align} \label{eq:invariantformulas}
\end{subequations}
The reason we choose to work with the particular choice of invariants defined via \cref{eq:the_alephs} is to reduce the degree of the polynomial that appears in the numerator of the invariant when expressed in powers of $\xi$, as can be observed by comparing the explicit expressions in \cref{eq:invariantformulas} with \cref{eq:tomo2d}. This leads to simplifications when performing numerical computations discussed below.

Now, to perform brane tomography, we plot the contours in the $(y_{\text{brane}}, \alpha)$-plane that correspond to the values of the invariants $(\aleph_1, \aleph_2)$ extracted from the entanglement velocity. The actual values of $(y_{\text{brane}}, \alpha)$ realized in the microstate geometry correspond to the intersection point of the constant $\aleph_1, \aleph_2$ contours, provided it lies within the physically permissible domain of the brane parameters. For the brane location, this corresponds to $-\pi/2 < y_{\text{brane}} \leq 0$. For the JT coupling $\alpha$, there are two possibilities, as discussed in detail in \cref{sec:DGP_bounds_subcrit}. If one only considers the requirement that the entanglement entropy does not exhibit ultrafast growth, \cref{eq:ve_bound_2d}, to be consistent with the physical picture of entanglement propagating via local interactions in the boundary theory, the physically permissible values of $\alpha$ satisfy the weak bound $|\alpha|\left(1+\sin y_\brane\right) \le 1$, c.f., \cref{eq:weak_bound_2d}. On the other hand, if one is willing to consider RT surfaces which pass through the singularity, \cref{eq:alpha_bound}, we have the much stricter constraint $|\alpha|\left(1-\sin y_\brane\right) \le 1$.

It turns out that only a small subset of the parameter space for the invariants $(\aleph_1, \aleph_2)$ translates into physical values for the brane parameters. 
If a particular set of $(\aleph_1, \aleph_2)$ does not correspond to physical brane parameters, it implies that the approach of modeling the CFT state holographically via a planar black hole with an ETW brane having JT gravity localized on it requires modifications. For instance, such a scenario might imply that there are additional couplings present on the ETW brane, beyond the JT term.

By numerically exploring a large subset of the space of constant $\aleph_1, \aleph_2$ contours, we find that the curves seem to intersect at most at one point in the entire $(y_{\text{brane}}, \alpha)$-plane. The fact that there is at most one intersection point over a large region in the parameter space of the invariants makes brane tomography sound plausible. Once we numerically locate the intersection point in the $(y_\brane, \alpha)$ plane for a given set of values for the invariants $(\aleph_1, \aleph_2)$, we can easily read off the values for the brane parameters.

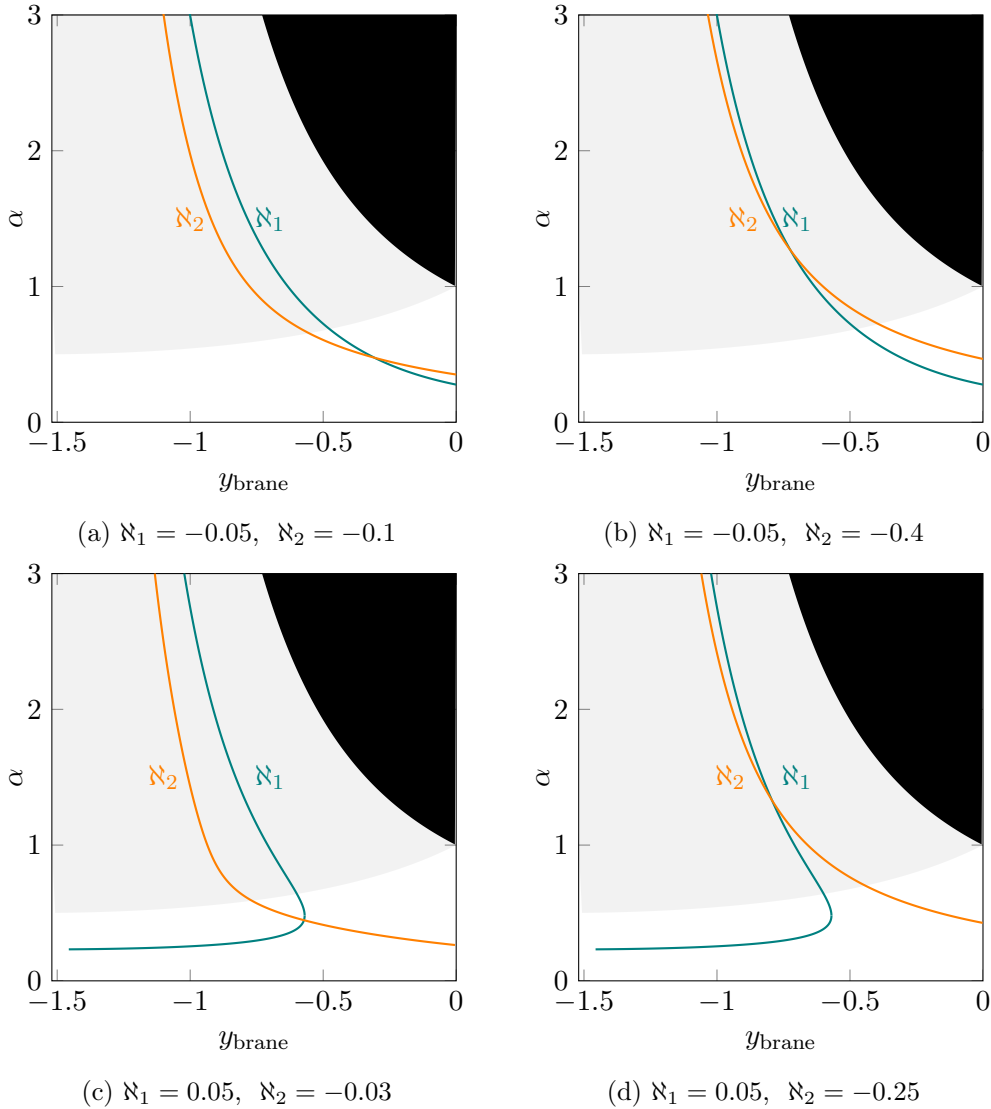
\begin{figure}[t]
    \centering
    \subcaptionbox{$\aleph_1 = -0.05$, \ $\aleph_2 = -0.1$}[0.45\textwidth]{
    \begin{tikzpicture}
    \begin{axis}[xlabel=$y_\brane$,ylabel=$\alpha$, ylabel shift = -0.2cm, xmin=-1.52, xmax=0, ymin=0, ymax=3, x=3.5cm, y=1.8cm, legend pos=south west, legend cell align={left}]
    \addplot[draw=none, name path=weak] file {tomodata/weakbound.dat};
    \addplot[draw=none, name path=strong] file {tomodata/strongbound.dat};
    \path[name path=axis] (axis cs:-1.57,3) -- (axis cs:0,3);
    \addplot [fill=gray, opacity=0.1] fill between[of=strong and axis];
    \addplot [fill=black] fill between[of=weak and axis];
    \addplot [thick, teal] file {tomodata/tomoNNal1.dat};
    \addplot [thick, orange] file {tomodata/tomoNNal2.dat};
    \node[teal] at (axis cs: -0.7,1.5) {$\aleph_1$};
    \node[orange] at (axis cs: -1.,1.5) {$\aleph_2$};
    \end{axis}
    \end{tikzpicture}}
    \subcaptionbox{$\aleph_1 = -0.05$, \ $\aleph_2 = -0.4$}[0.45\textwidth]{
    \begin{tikzpicture}
    \begin{axis}[xlabel=$y_\brane$,ylabel=$\alpha$, ylabel shift = -0.2cm, xmin=-1.52, xmax=0, ymin=0, ymax=3, x=3.5cm, y=1.8cm, legend pos=south west, legend cell align={left}]
    \addplot[draw=none, name path=weak] file {tomodata/weakbound.dat};
    \addplot[draw=none, name path=strong] file {tomodata/strongbound.dat};
    \path[name path=axis] (axis cs:-1.57,3) -- (axis cs:0,3);
    \addplot [fill=gray, opacity=0.1] fill between[of=strong and axis];
    \addplot [fill=black] fill between[of=weak and axis];
    \addplot [thick, teal] file {tomodata/tomoNNal1out.dat};
    \addplot [thick, orange] file {tomodata/tomoNNal2out.dat};
    \node[teal] at (axis cs: -0.7,1.5) {$\aleph_1$};
    \node[orange] at (axis cs: -0.9,1.5) {$\aleph_2$};
    \end{axis}
    \end{tikzpicture}}
    \subcaptionbox{$\aleph_1 = 0.05$, \ $\aleph_2 = -0.03$}[0.45\textwidth]{
    \begin{tikzpicture}
    \begin{axis}[xlabel=$y_\brane$,ylabel=$\alpha$, ylabel shift = -0.2cm, xmin=-1.52, xmax=0, ymin=0, ymax=3, x=3.5cm, y=1.8cm, legend pos=south west, legend cell align={left}]
    \addplot[draw=none, name path=weak] file {tomodata/weakbound.dat};
    \addplot[draw=none, name path=strong] file {tomodata/strongbound.dat};
    \path[name path=axis] (axis cs:-1.57,3) -- (axis cs:0,3);
    \addplot [fill=gray, opacity=0.1] fill between[of=strong and axis];
    \addplot [fill=black] fill between[of=weak and axis];
    \addplot [thick, teal] file {tomodata/tomoPNal1one.dat};
    \addplot [thick, teal] file {tomodata/tomoPNal1two.dat};
    \addplot [thick, orange] file {tomodata/tomoPNal2.dat};
    \node[teal] at (axis cs: -0.7,1.5) {$\aleph_1$};
    \node[orange] at (axis cs: -1.1,1.5) {$\aleph_2$};
    \end{axis}
    \end{tikzpicture}}
    \subcaptionbox{$\aleph_1 = 0.05$, \ $\aleph_2 = -0.25$}[0.45\textwidth]{
    \begin{tikzpicture}
    \begin{axis}[xlabel=$y_\brane$,ylabel=$\alpha$, ylabel shift = -0.2cm, xmin=-1.52, xmax=0, ymin=0, ymax=3, x=3.5cm, y=1.8cm, legend pos=south west, legend cell align={left}]
    \addplot[draw=none, name path=weak] file {tomodata/weakbound.dat};
    \addplot[draw=none, name path=strong] file {tomodata/strongbound.dat};
    \path[name path=axis] (axis cs:-1.57,3) -- (axis cs:0,3);
    \addplot [fill=gray, opacity=0.1] fill between[of=strong and axis];
    \addplot [fill=black] fill between[of=weak and axis];
    \addplot [thick, teal] file {tomodata/tomoPNal1oneout.dat};
    \addplot [thick, teal] file {tomodata/tomoPNal1twoout.dat};
    \addplot [thick, orange] file {tomodata/tomoPNal2out.dat};
    \node[teal] at (axis cs: -0.7,1.5) {$\aleph_1$};
    \node[orange] at (axis cs: -0.95,1.5) {$\aleph_2$};
    \end{axis}
    \end{tikzpicture}}
    \caption{Contours corresponding to constant values for the invariants $\aleph_1$ (blue) and $\aleph_2$ (orange) in the $(y_{\text{brane}}, \alpha)$-plane. If one imposes the stronger (weaker) bound on the JT coupling $\alpha$, then the physically permissible domain of the brane parameters is given by the white (white + grey) region. One can easily read off the microstate parameters $(y_{\text{brane}}, \alpha)$ from the location of the intersection point.
    }
    \label{fig:tomography}
\end{figure}

In \cref{fig:tomography}, as an illustrative example of performing tomography when $\alpha > 0$, we plot several constant $\aleph_{1}, \aleph_2$ contours on the $(y_{\text{brane}}, \alpha)$-plane. Each pair of curves represents a point in the parameter space of the invariants $(\aleph_1,\aleph_2)$. The intersection points correspond to the values of $y_\brane, \alpha$ realized in the bulk geometry.

Now, assuming that one has deduced the values of the parameters $(y_\brane, \alpha)$ using the protocol outlined above, one can proceed to obtain the parameter $t_0$. Recall that the expansion in \cref{eq:asymptopia} was obtained assuming $t_0 = 0$. Now, for the tomographically extracted values of $(y_\brane,\alpha)$, one can compute the value for the coefficient $\mathcal{K}_1$, \cref{eq:tomo2d1}. If this value matches with the value that actually appears in the asymptotic expansion \cref{eq:asymptopia}, then one can conclude that the bulk geometry is the one where $t_0 = 0$. However, if the two values differ, then knowing that the difference can be accounted for by a multiplicative factor of $e^{-\frac{4\pi t_0}{\beta}}$, one can compute the non-zero value for $t_0$.

These three parameters fix the subregion entropy for the boundary CFT almost completely as a function of time. The only missing piece is the constant term in \cref{eq:sint_full_2d_2}, which can be determined by computing the relative entropy between our state and the thermal state, \cref{eq:rel_entropy}, at an arbitrary instant in time, and comparing the result with the one that follows by using the tomographically determined brane parameters.
It is advantageous to make use of the relative entropy as the dependence of the entanglement entropy on the choice of the cutoff scale $\epsilon$ cancels out automatically.

One may also wonder about the utility of higher order invariants that can be constructed by making use of even more subleading terms in the asymptotic expansion \cref{eq:asymptopia}, beyond the first three which we used to construct $\aleph_1$ and $\aleph_2$. Such invariants in principle overdetermine the tomography protocol outlined above and therefore provide an important consistency check on our model. Given a set of parameters $(y_\brane, \alpha)$ extracted using tomography, one can compute the value for a particular higher-order invariant using its functional form. If the CFT state is correctly described by an ETW brane with the given parameters, then the computed value for the higher-order invariant must agree with its value directly obtained from the asymptotic behaviour of the entanglement velocity. On the other hand, in case of a disagreement, we can conclude that modelling the CFT state using an ETW brane with JT gravity localized on it is not sufficient.

As an example, a more general model (but still one with an ETW brane) is obtained by adding interaction terms for the dilaton $\varphi$, which naturally arise when considering more general two-dimensional dilaton gravity models \cite{Grumiller:2021cwg}. For such cases, one has to consider invariants beyond the two we considered above, and employ tomography to deduce information about the additional couplings. This will of course entail performing numerical calculations in a multidimensional space, beyond the simple two-dimensional plots of \cref{fig:tomography}, with additional axes denoting new brane parameters of interest.\footnote{Another set of nontrivial terms that can be present in the brane action, \cref{eq:braneaction}, are curvature invariants constructed out of the extrinsic curvature tensor $K_{\mu\nu}$, beyond the leading Gibbons-Hawking-York term. However, in \cref{sec:higher_curv}, we argue that such terms do not affect the dynamics of the brane, at least for corrections which are quadratic in $K_{\mu\nu}$.}

\subsection{Critical brane tomography}
The discussion in the previous subsection was only concerned with the subcritical case, but one can of course utilize brane tomography for the critical case as well. One can simply repeat the same steps as before, with slight modifications. One first determines the sign of $\alpha$ and derives the coefficients \cref{eq:tomo2d} in the asymptotic expansion of $v_E$. Now, however, there is no parameter $y_\brane$ anymore, since our brane is located at the limiting point $y_\brane = - \frac{\pi}{2}$. This simplifies the problem, and the coefficients in the asymptotic expansion only depend upon the single parameter $\alpha_\text{cr}$, which can be obtained by forming an invariant out of $\mathcal K_1$ and $\mathcal K_2$. Subsequently, given the value of $\alpha_\text{cr}$, one can determine the parameter $t_0$ in the solution, as well as the constant piece of the dilaton, following the same procedure as discussed above for the subcritical case.

Instead of obtaining the coefficients $\mathcal K_i$ from scratch, it is instructive to derive them using the relation between the couplings in the critical and subcritical case, \cref{eq:relation_alpha}. As this equation implies, the coupling $\alpha$ for subcritical brane tomography has to be taken to infinity to obtain a finite $\alpha_\text{cr}$, and this might seem problematic at first sight. Even worse, naively substituting \cref{eq:relation_alpha} into the expressions for $\mathcal K_i$ in the subcritical case, \cref{eq:tomo2d}, one finds that all the coefficients seem to vanish. However, note that in obtaining the critical case as a limit from the subcritical case, we also had to perform a time-translation, which diverged as we sent the brane location $y_\brane \to - \frac \pi 2$, c.f., \cref{eq:fixed}. Accounting for this, each exponent in \cref{eq:asymptopia} comes with a diverging factor, which precisely cancels the factor that made the $\mathcal K_i$ vanish.

Of course, none of these subtleties affect the invariants, since they are defined to be invariant under time translations. Therefore, in order to obtain the correct expressions for the invariants we only need to substitute \cref{eq:relation_alpha} and take $y_\brane \to - \frac \pi 2$ in \cref{eq:invariantformulas} to get
\begin{subequations}
\begin{align}
\label{eq:alephs_crit_1}
    \aleph_1 = \frac{2 \alpha_\text{cr}-2}{(\alpha_\text{cr}-2)^2},\\
    \label{eq:alephs_crit_2}
    \aleph_2 = \frac{3 \alpha_\text{cr}-2}{(\alpha_\text{cr}-2)^3}.
\end{align}
\end{subequations}
At least for generic points, there is no $\alpha_\text{cr}$ such that \cref{eq:alephs_crit_1,eq:alephs_crit_2} agree with \cref{eq:def_aleph_1,eq:def_aleph_2}. Thus, not only can we perform tomography for the critical case as well, we can also tell apart the critical and subcritical cases.

\section{Discussion and outlook}
\label{sec:discussion}
In this paper we have discussed how bounds on the entanglement velocity for a holographic CFT can bound gravitational brane couplings, if the dual gravitational description is provided by black hole microstates with an end-of-the-world brane behind the horizon. We also discussed a protocol to perform brane tomography based on examining the late time behaviour of entanglement growth. We focused on the case of AdS${}_3$ planar black hole geometries with ETW branes behind their horizons. This case is particularly attractive, as many of the details can be worked out analytically. Nonetheless, the general strategy is applicable to other, higher dimensional situations as well.
For our analysis, we have operated under the implicit assumption that the RT prescription is valid throughout the bulk geometry. It is of course possible that this effective description for a high energy pure state of a CFT, in terms of black hole geometries with ETW branes is only valid for a certain spacetime region in the bulk. It would be interesting to have an independent argument for when this could happen, in order to further refine the bounds we have computed.

Our bounds on the JT coupling rely on the bound on entropy growth in the dual CFT state. Interestingly, the issues which can arise in the growth rate of entanglement entropy, which we used to derive our bounds, are unrelated to another suspicious feature which arises for negative DGP couplings in models of evaporating black holes. As discussed in appendix B of \cite{Chen:2020uac}, for negative DGP coupling, the so-called RT-bubbles appear, which are self-supporting RT surfaces with a negative generalized entropy and are homologous to the empty set. However, they appear for arbitrarily small-but-negative DGP couplings, and are therefore not ruled out by our results.

Let us now make a few comments on some possible extensions of the present analysis. Investigating quantities which are affected by the JT coupling, other than the holographic entanglement entropy, might allow one to tighten the bounds provided in this paper, or derive new bounds altogether. A good target would be bounds on the constant piece of the dilaton, which remains unconstrained in our analysis. In the context of entanglement velocity, the recent paper \cite{Mezei:2019sla} provided an even tighter bound at early times than the one we have utilized, which however does not seem to give new information for the case at hand. Another interesting option would be to try to understand whether the linear growth of complexity can provide tighter bounds for our setup \cite{Hernandez:2020nem}.

For the present analysis, we restricted our attention to the possibility of JT gravity on the brane. The most general consistent deformation of JT gravity is given by \emph{generalized dilaton gravity} models in two dimensions \cite{Grumiller:2021cwg}, with an action of the form
\begin{equation*}
    I_{DG} = \frac{1}{16\pi G_N^\brane} \int d^2x \sqrt{-h} \left(\varphi R^\brane - 2 V\!\!\left(\varphi, - (\partial\varphi)^2\right)\right).
\end{equation*}
Here, the potential $V$ is an arbitrary function of the dilaton and its kinetic term.\footnote{JT
gravity appears as a special case of this action, with $V = \varphi \Lambda^\brane$.} Depending upon the choice for the potential $V$, adding this term to the brane action can give rise to several additional couplings, which will satisfy their own bounds that will follow by generalizing our analysis. With additional couplings present, performing brane tomography will be much more challenging as well. Of course, there is also the possibility of additional matter fields living on the brane, further enlarging the landscape of possible couplings.\footnote{See for instance \cite{Moitra:2019xoj, Banks:2022pnc} for explorations of JT gravity coupled to matter fields.}

Another obvious generalization of the above analysis pertains to the study of the higher dimensional case.\footnote{The DGP term in higher dimensions can also be augmented with additional higher curvature correction terms constructed out of the Riemann tensor for the induced metric on the brane.} For such cases, two important issues arise. The first is that higher dimensional generalizations are much less tractable analytically, and one generally has to resort to a numerical approach. More problematic, however, is the issue that the derivation of the bound on the entanglement velocity, reviewed in \cref{sec:limits_on_ve}, required the explicit form of the modular Hamiltonian of an interval for the two-dimensional CFT thermal state. The natural generalization to higher dimensions will involve the thermal modular Hamiltonian on a strip, whose explicit form is not known. In the absence of sufficient knowledge about the form of the modular Hamiltonian in higher dimensions, it is not obvious that bounds on entanglement growth such as \cref{eq:ve_bound_2d} still hold true. One way out is to look for other quantities for the CFT state that are bounded, and study how these could possibly translate into bounds on the couplings. We hope to report on results for  higher dimensional generalizations in the future \cite{toappear}.

In a particular limit, ETW branes give rise to double-holography, i.e., an effective holographic description of the bulk spacetime on the asymptotic boundary together with the brane. In the limit where the brane sits close to the left asymptotic boundary, it is generally expected that the theory on the brane is described by an effective quantum field theory coupled to gravity, plus higher order curvature corrections. It would be interesting to understand --- and perhaps even prove --- our bounds from the point of view of this effective description. In particular, it would be interesting to see whether or not our strong bound on the coupling can be arrived at from a doubly-holographic perspective. If the answer turns out to be yes, this would suggest that the extremal surfaces passing through the singularity should indeed be taken seriously.

Our results provide us with two-dimensional ETW brane solutions for which RT surfaces can end at various locations on the brane. In the double-holography description, the region bounded by the RT surface on the brane is a quantum extremal island. Thus, our solutions complement the analyses of \cite{Hartman:2020khs,Bousso:2022gth} by providing solutions where islands appear in more general locations. This poses the question whether the presence of islands in a flat, radiation dominated universe is only a property of JT gravity, or is also true in other dimensions in gravitational theories with higher curvature corrections.

Furthermore, in higher dimensions, it was proposed recently in \cite{Waddell:2022fbn} to introduce additional defect branes in the bulk. Additional branes, or bulk defects, will affect the analysis performed here, and it would be interesting to investigate whether this leads to the bounds we find becoming stronger or weaker. This is important insofar as the analysis of \cite{Waddell:2022fbn} found that in order to realize branes which localize gravity using a Euclidean path integral, the defect branes are not sufficient by themselves, and that negative gravitational couplings also need to be added to the brane.

Lastly, another interesting direction to explore is to what extent the results presented here carry over to the case of black holes with spherical horizons \cite{Cooper:2018cmb}. For such cases, the size of the accessible boundary region in the CFT is bounded from above, thereby preventing us from choosing an arbitrarily large interval on the boundary. This of course does not immediately affect the bounds on entanglement velocity. However, for particular parameters the connected extremal surfaces might never be the smallest ones and thus one would never have the situation that the entanglement velocity violates any bound. Moreover, spherical horizon geometries would jeopardize the tomographic analysis of \cref{sec:tomography}, since the asymptotic behaviour of entanglement growth is not accessible anymore. In that situation, one might still be able to obtain information about the brane parameters by focusing on the time-dependence of the entanglement entropy around its minimum.

\acknowledgments
We thank Tom Hartman, Mark Mezei, Ayan Mukhopadhyay, Rob Myers, Brian Swingle and Chris Waddell for helpful comments and discussions. JHL is supported by the Perimeter Institute for Theoretical Physics and in part by the NSERC Discovery Grant program. DN thanks the Simons Foundation for support through the \emph{It from Qubit} collaboration. The work of AS is also supported in part by the European Research Council (ERC) under the European Union’s Horizon 2020 research and innovation program (grant agreement no.\ 758759). Research at Perimeter Institute is supported in part by the Government of Canada through the Department of Innovation, Science and Economic Development Canada and by the Province of Ontario through the Ministry of Colleges and Universities.

\appendix

\section{Brane dynamics}
\label{app:brane_dynamics}
In this appendix, we present the derivation of \cref{eq:var_tot,eq:brane_eom} that govern the dynamics of the ETW brane. We also discuss the possibility of modifying the brane action \cref{eq:braneaction} with higher extrinsic curvature correction terms, and their resulting effects on the dynamics. For the sake of generality, in this appendix, we present our calculations for a $d+1$ dimensional spacetime, $d \ge 2$. The bulk, brane and boundary actions are thus given by
\begin{subequations}
\begin{align}
 I_\bulk &= \frac{1}{16\pi G_N} \int d^{d+1}x \sqrt{-g} \left(R -2 \Lambda \right), \label{eq:Ibulkgen}\\
 I_\brane &= \frac{1}{8\pi G_N} \int d^dy \sqrt{-h} \left(K - \frac{(d-1)}{L} \, T_0 \right), \label{eq:Ibranegen} \\
 I_\bdry &= \frac 1 {8 \pi G_N} \int d^dy \sqrt{-h} \, K + \text{(counterterms)},\label{eq:Ibdrygen}
 \end{align}
\end{subequations}
where $\Lambda = - \frac{d(d-1)}{2L^2}$ is the cosmological constant. For notational convenience, we denote the bulk coordinates via $x^\mu$, $\mu = 0,1, \ldots d$, and the brane/boundary coordinates via $y^a$, $a = 0,1, \ldots d-1$. The embedding of the brane in the bulk can be thought of as parametrized by $x^\mu \equiv x^\mu(y^a)$. The tangent vectors to the brane are then given by $e^\mu_a = \frac{\partial x^\mu}{\partial y^a}\big|_{\text{brane}}$. The induced metric on the brane $h_{ab}$ is related to the bulk metric via the projection $h_{ab} = e^\mu_a e^\nu_b g_{\mu\nu}$, with $h$ being its determinant.

\subsection{The brane trajectory}
\label{sec:brane_eom}
Under a variation of the bulk metric $g_{\mu\nu}$, the variation of the bulk action \cref{eq:Ibulkgen} is
\begin{equation}
\delta I_{\textrm{bulk}} = \frac{1}{16\pi G_N} \int d^{d+1}x \sqrt{-g} \, G_{\mu\nu} \, \delta g^{\mu\nu} + \frac{1}{16\pi G_N} \int_{\textrm{brane}} d^{d}y \sqrt{-h}\, n^\mu h^{\alpha\beta} \left( \nabla_\alpha\delta g_{\mu\beta} - \nabla_{\mu} \delta g_{\alpha\beta}\right).
\label{eq:bulk_var}
\end{equation}
Here $G_{\mu\nu} \equiv R_{\mu\nu} - \frac{1}{2} g_{\mu\nu} R + \Lambda g_{\mu\nu}$ is the Einstein tensor in the bulk.  $n^\mu$ denotes the unit spacelike normal to the brane, satisfying $n^\mu n_\mu = 1$.\footnote{Given the brane trajectory $f(x^\mu) = 0$, the unit spacelike normal to it is $n_\mu = \frac{\partial_\mu f}{\sqrt{g^{\alpha\beta} \partial_\alpha f \partial_\beta f}}.$} The quantity $h_{\mu\nu}$ appearing in $\delta I_\bulk$ is the transverse metric or the projector, defined via $h_{\mu\nu} = g_{\mu\nu} - n_{\mu} n_{\nu}$, and thus satisfies $n^\mu h_{\mu\nu} = 0$. The first term in $\delta I_\bulk$ vanishes onshell. The second term is a combination of two terms, one proportional to the tangential derivative of the metric variation, $h^{\alpha\beta} \nabla_\alpha\delta g_{\mu\beta}$, and the other proportional to the normal derivative of the metric variation, $n^\mu \nabla_\mu \delta g_{\alpha\beta}$.

Note that apart from the terms presented in \cref{eq:bulk_var}, the variation of the bulk action also gives rise to terms on the asymptotic AdS boundary. These cancel off with the Gibbons-Hawking term \cref{eq:Ibdrygen}, assuming that one imposes Dirichlet boundary conditions for the metric variation $\delta g_{\mu\nu}$ at the AdS boundary, which we indeed do. 

The variation of the brane action \cref{eq:Ibranegen} gives
\begin{equation}
\label{eq:brane_var}
\begin{split}
 &\delta I_{\textrm{brane}} = 
 \frac{1}{16\pi G_N} \int_{\textrm{brane}} d^d y \sqrt{-h} \left(2 K_{\mu\nu} -  g_{\mu\nu} K + \frac{d-1}{L} \, T_0 h_{\mu\nu}\right) \delta g^{\mu\nu} \\
 &\qquad\qquad - \frac{1}{16\pi G_N} \int_{\textrm{brane}} d^d y \sqrt{-h} \, n^\mu h^{\alpha\beta} \left(2 \nabla_\alpha \delta g_{\mu\beta} - \nabla_\mu \delta g_{\alpha\beta} \right) .
 \end{split}
\end{equation}
In obtaining this result we have made use of
\begin{subequations}
\begin{align}
    &\delta n_\mu = - \frac{1}{2} n_\mu n_\rho n_\sigma \delta g^{\rho\sigma}\, ,\\
    &\delta n^{\mu} = n_\nu \delta g^{\mu\nu} - \frac{1}{2} n^\mu n_\rho n_\sigma \delta g^{\rho\sigma}\, ,\\
    &\delta K = K_{\mu\nu} \delta g^{\mu\nu} - \frac{1}{2} K n_\mu n_\nu \delta g^{\mu\nu} - \frac{1}{2} n^\mu h^{\alpha\beta} \left( \nabla_\alpha \delta g_{\mu\beta} + \nabla_\beta \delta g_{\mu\alpha} - \nabla_\mu \delta g_{\alpha\beta}\right), \label{eq:varK}
\end{align}
\end{subequations}
which are straightforward to obtain.

Finally, combining \cref{eq:bulk_var,eq:brane_var} gives us the total variation to be
\begin{equation}
\begin{split}
    &\delta I_\total = \frac{1}{16\pi G_N} \int d^{d+1}x \sqrt{-g} \, G_{\mu\nu} \delta g^{\mu\nu} \\ 
    &\qquad\quad\, + \frac{1}{16 \pi G_N} \int_{\textrm{brane}} d^d y \sqrt{-h} \left( 2 K_{\mu\nu} - g_{\mu\nu} K + \frac{d-1}{L}\, T_0 h_{\mu\nu}\right) \delta g^{\mu\nu} \\ 
    &\qquad\quad\, - \frac{1}{16 \pi G_N} \int_{\textrm{brane}} d^d y \sqrt{-h} \, n^\mu h^{\alpha\beta} \nabla_\alpha\delta g_{\mu\beta}.
    \label{eq:delta_I}
\end{split}
\end{equation}
We can further simplify the last term of eq.\ \eqref{eq:delta_I}, which is proportional to the tangential derivative of the metric variation. Simple algebraic manipulations yield
\begin{align}
     &- \frac{1}{16 \pi G_N} \int_{\textrm{brane}} d^d y \sqrt{-h} \, n^\mu h^{\alpha\beta} \nabla_\alpha\delta g_{\mu\beta} = - \frac{1}{16 \pi G_N}  \int_{\textrm{brane}} d^d y \sqrt{-h} \left( K_{\mu\nu} - K n_\mu n_\nu\right)\delta g^{\mu\nu} \nonumber\\
     &\qquad\qquad\qquad\qquad\qquad - \frac{1}{16 \pi G_N}  \int_{\textrm{brane}} d^d y \sqrt{-h} \left[\nabla_\alpha\left(n^\mu h^{\alpha\beta} \delta  g_{\mu\beta}\right) + n^\mu n^\alpha \nabla_\alpha n^\beta \delta g_{\mu\beta}\right].\label{eq:furt_simp}
\end{align}
Notice that the vector $V^\alpha \equiv n^\mu h^{\alpha\beta}\delta g_{\mu\beta}$ is tangential to the brane, since $n_\alpha V^\alpha = 0$. We can define the projection of its covariant derivative along the brane as $\bar{\nabla}_\alpha V_\beta \equiv h^\mu_\alpha h^\nu_\beta \nabla_\mu V_\nu$. Contracting both sides by the projector $h^{\alpha\beta} \equiv e^\alpha_a e^\beta_b h^{ab}$ yields
\begin{equation}
    h^{ab}\left(e^\alpha_a e^\beta_b \bar{\nabla}_\alpha V_\beta\right) = \nabla_\alpha V^\alpha - n^\alpha n_\beta \nabla_\alpha V^\beta. \label{eq:proj_cov_1}
\end{equation}
The quantity in the parentheses on the left is the pull-back of the projected covariant derivative $\bar{\nabla}_\alpha V_\beta$ to the brane, i.e., $e^\alpha_a e^\beta_b \bar{\nabla}_\alpha V_\beta = \nabla^{(h)}_a v_b$, where $\nabla^{(h)}_a$ is the covariant derivative on the brane defined with respect to the induced metric $h_{ab}$, and $v_a = e^\alpha_a V_\alpha$ is the pull-back of the vector $V_\alpha$.  With this, \cref{eq:proj_cov_1} becomes
\begin{equation}
    \nabla_\alpha\left(n^\mu h^{\alpha\beta} \delta  g_{\mu\beta}\right) + n^\mu n^\alpha \nabla_\alpha n^\beta \delta g_{\mu\beta} = \nabla^{(h)}_a v^a.\label{eq:proj_cov_2}
\end{equation}
Thus, the last term in \cref{eq:furt_simp} is the integral of a total divergence,
\begin{equation}
    \int_{\textrm{brane}} d^d y \sqrt{-h} \left[\nabla_\alpha\left(n^\mu h^{\alpha\beta} \delta  g_{\mu\beta}\right) + n^\mu n^\alpha \nabla_\alpha n^\beta \delta g_{\mu\beta}\right] = \int_{\textrm{brane}} d^d y \sqrt{-h} \, \nabla^{(h)}_a v^a,\label{eq:tot_deriv}
\end{equation}
which vanishes. Using the result \cref{eq:tot_deriv} in \cref{eq:furt_simp} and combining with \cref{eq:delta_I} yields 
\begin{equation}
\begin{split}
    &\delta I_\total = \frac{1}{16\pi G_N} \int d^{d+1}x \sqrt{-g} \, G_{\mu\nu} \delta g^{\mu\nu} \\ 
    &\qquad\quad\, + \frac{1}{16 \pi G_N} \int_{\textrm{brane}} d^d y \sqrt{-h} \left( K_{\mu\nu} - h_{\mu\nu} K + \frac{d-1}{L}\, T_0 h_{\mu\nu}\right) \delta g^{\mu\nu}.
    \label{eq:delta_I_fin}
\end{split}
\end{equation}
This is the result presented in \cref{eq:var_tot} of the main text, with $d=2$. The quantity inside the parentheses must vanish for the total variation $\delta I_\total$ to vanish onshell, since we do not impose the Dirichlet boundary condition on the metric variation $\delta g^{\mu\nu}$ at the location of the brane. This gives 
\begin{equation}
    K_{ij} - K h_{ij} + \frac{d-1}{L}\, T_0 h_{ij} = 0,
    \label{eq:brane_traj}
\end{equation}
where we have performed a pull-back. \Cref{eq:brane_traj} determines the trajectory of the brane. Further taking its trace with respect to the induced metric $h_{ij}$ gives $K = T_0 d/L$, which when substituted back into \cref{eq:brane_traj} yields \cref{eq:brane_eom} of the main text.

\subsection{Modifying the brane action}
\label{sec:higher_curv}
We now consider the possibility of including higher (extrinsic) curvature correction terms to the ETW brane action eq.\ \eqref{eq:Ibranegen}, and how these might impact the resulting brane dynamics. Consider the following higher extrinsic curvature correction term, denoted by $\tilde{I}$, added to the brane action eq.\ \eqref{eq:Ibranegen},
\begin{equation}
    \label{eq:brane_corr}
    \tilde{I}_{\text{brane}} = \frac{1}{8\pi G_N} \int_{\text{brane}} d^d y \sqrt{-h} \left(\lambda_1 K^{\mu\nu} K_{\mu\nu} + \lambda_2 K^2 \right).
\end{equation} 
Here $\lambda_1, \lambda_2$ are constants. The variation of $\tilde{I}$ under the variation of the metric gives
\begin{align}
    &\delta \tilde{I}_{\text{brane}} = - \frac{1}{16\pi G_N} \int_{\text{brane}} d^d y \sqrt{-h} \, h_{\alpha\beta} \, \delta g^{\alpha\beta} \left( \lambda_1 K^{\mu\nu} K_{\mu\nu} + \lambda_2 K^2\right) \nonumber\\
    &+ \frac{1}{4\pi G_N} \int_{\text{brane}} d^d y \sqrt{-h} \bigg[\frac{1}{2} n^\mu \left(\lambda_1 K^{\alpha\beta} + \lambda_2 K h^{\alpha\beta}\right)  \left(\nabla_\mu \delta g_{\alpha\beta} - \nabla_\alpha \delta g_{\mu\beta} - \nabla_\beta \delta g_{\mu\alpha} \right)  \nonumber\\
    &\qquad\qquad - \frac{1}{2} \left( \lambda_1 K^{\mu\nu} K_{\mu\nu} + \lambda_2 K^2 \right) n_\alpha n_\beta \, \delta g^{\alpha\beta} + \lambda_1 K_{\mu\nu} K^{\nu}_{\rho} \delta g^{\mu\rho} + \lambda_2 K K_{\mu\nu} \delta g^{\mu\nu} \bigg], \label{eq:high_var}
\end{align}
which is straightforward to obtain making use of \cref{eq:varK} and 
\begin{equation}
\begin{split}
    &\delta K_{\mu\nu} = - \frac{1}{2} K_{\mu\nu} n_\rho n_\sigma \delta g^{\rho\sigma} - \left( n_\mu K_{\rho\nu} + n_\nu K_{\rho\mu}\right) n_\sigma \delta g^{\rho\sigma} \\
    &\qquad\quad\, - \frac{1}{2} h_\mu^\alpha h_\nu^\beta n^\rho \left( \nabla_\alpha \delta g_{\rho\beta} + \nabla_\beta \delta g_{\rho\alpha} - \nabla_\rho \delta g_{\alpha\beta}\right).
\end{split}
\end{equation}

The term proportional to the normal derivative of the metric variation, appearing in the second line of eq.\ \eqref{eq:high_var}, is worrisome. The only way to get rid of this term without imposing special boundary conditions is to assume that its coefficient vanishes identically, i.e.,
\begin{equation}
\label{eq:constraint_lamb}
    \lambda_1 K^{\alpha\beta} + \lambda_2 K h^{\alpha\beta} = 0.
\end{equation}
Taking trace of this condition with the metric $g_{\alpha\beta}$ gives us the following constraint on the possible values for the parameters $\lambda_1, \lambda_2$,
\begin{equation}
\label{eq:l1l2}
    \lambda_1 + d \lambda_2 = 0.
\end{equation}
Putting this back into eq.\ \eqref{eq:constraint_lamb} gives
\begin{equation}
\label{eq:constraint_lamb2}
    K_{\alpha\beta} - \frac{1}{d} \, K h_{\alpha\beta} = 0.
\end{equation}

With the higher curvature terms eq.\ \eqref{eq:brane_corr} taken into consideration along with the original brane action eq.\ \eqref{eq:Ibranegen}, the equation for the brane trajectory can be obtained by combining the results \cref{eq:delta_I_fin,eq:high_var}, and is given by
\begin{equation}
    \begin{split}
    K_{\alpha\beta} - K h_{\alpha\beta} + \frac{d-1}{L}\, T_0 h_{\alpha\beta} - \left(h_{\alpha\beta} + 2n_\alpha n_\beta\right)\left(\lambda_1 K^{\mu\nu} K_{\mu\nu} + \lambda_2 K^2\right)& \\
    + 4\left(\lambda_1 K_{\alpha\rho} K^\rho_\beta + \lambda_2 K K_{\alpha\beta}\right)& = 0,
    \end{split}
\end{equation}
where we have made use of the constraint eq.\ \eqref{eq:constraint_lamb} to kill the terms proportional to the derivatives of the metric variation appearing in eq.\ \eqref{eq:high_var}. Interestingly, further imposing eqs.\ \eqref{eq:l1l2} and \eqref{eq:constraint_lamb2}, which are consequences of the constraint eq.\ \eqref{eq:constraint_lamb}, kills all the additional terms that arise because of the higher curvature corrections, and we are once again left with the original brane equation of motion, given by
\begin{equation}
    K_{\alpha\beta} - K h_{\alpha\beta} + \frac{d-1}{L}\, T_0 h_{\alpha\beta} = 0.
\end{equation}
This shows that it is not very fruitful to consider higher extrinsic curvature correction terms to the brane action \cref{eq:braneaction}, at least to the leading order, \cref{eq:brane_corr}.

\section{Extremal surfaces crossing between universes}
\label{app:universe_crossing}
In this appendix, we give a more detailed account of the extremal surfaces which cross between different universes i.e. pass through the singularity, c.f., \cref{fig:rt_other_universes}. Since these surfaces cross through a coordinate singularity, their extremality as well as existence is not clear immediately. 

To understand extremal surfaces which go between universes, it is useful to describe them in global AdS spacetime. Since AdS${}_3$ does not have any intrinsic gravitational dynamics, various regions covered by different coordinate systems are simply quotients or patches of the global AdS${}_3$ geometry. This of course is also true for the planar black hole geometry discussed in \cref{sec:setup}, which is simply the AdS$_3$ Rindler patch. In order to go between global coordinates and Kruskal-Szekeres coordinates $(\tau,y)$, we use the embedding of AdS${}_3$ in $\mathbb R^{2,2}$ given by
\begin{align}
    - X_0^2 + X_1^2 + X_2^2 - X_3^2 = -L^2.
\end{align}
Global coordinates parametrize this hypersurface via
\begin{align}
\begin{split}
    X_0 &= L \sin t \cosh \rho,\\
    X_1 &= L \cos \phi \sinh \rho,\\
    X_2 &= L \sin \phi \sinh \rho, \\
    X_3 &= L \cos t \cosh \rho.
\end{split}
\end{align}
On the other hand, our Kruskal-Szekeres coordinates $(\tau,y)$ parametrize only part of the hypersurface via
\begin{align}
\begin{split}
\label{eq:ks_parametrization}
    X_0 &= L \frac{\sin \tau}{\cos y},\\
    X_1 &= L \tan y,\\
    X_2 &= L \frac{\cos \tau}{\cos y} \sinh \left( \frac{r_+}{L^2} x \right),\\
    X_3 &= L \frac{\cos \tau}{\cos y} \cosh \left( \frac{r_+}{L^2} x \right).
    \end{split}
\end{align}
The region parametrized by \cref{eq:ks_parametrization} can roughly be characterized as the complement of Rindler patches.
The extremal surfaces we are considering have a constant value of $x$, and can be continued to values of $\tau > \frac \pi 2$ and $\tau < - \frac \pi 2$. We will now discuss the case of a single extremal surface at $x = 0$, as well as the case of two extremal surfaces at two different values for $x$.

\subsection{A single extremal surface}
We will focus on the case of $x = 0$. From the embedding equations for $X_2$ this implies that in global coordinates the extremal surface is located at $\phi = 0$ or $\phi = \pi$ (or $\rho = 0$). The relation for $X_1$ then implies that it is located at $\phi = 0$ for $y >0$, $\phi = \pi$ for $y < 0$, and $\rho = 0$ for $y=0$. This further implies that
\begin{align}
    \label{eq:y_vs_rho}
    \tan y = \pm \sinh \rho,
\end{align}
With this, the equation for $X_0$ tells us that we can identify $t$ with $\tau$. Therefore, since both coordinate patches cover the trajectory of an extremal surface at $x = 0$, this surface also clearly exists in the global AdS spacetime. Extremality of the surface is guaranteed by the $\mathbb Z_2$ symmetry $\phi \to - \phi$ in global AdS.

\subsection{Multiple extremal surfaces}
Let us now turn to the situation where we have multiple extremal surfaces which leave the boundary at a fixed $\tau_\bdry$ but at different values of $x$. As discussed in \cref{sec:DGP2D}, all such surfaces fall into the coordinate singularity at $\tau = - \frac \pi 2$. Let us now focus closely on this location. $X_2$ and $X_3$ vanish at this point. This requires that in the bulk $\tau = \pm \frac \pi 2$. Together with the requirement for positivity of $X_0$, this implies that $t = \frac \pi 2$. As before, vanishing of $X_2$ implies \cref{eq:y_vs_rho}.

Extremal surfaces emanating at the same $\tau_\bdry$ will arrive at $\tau = - \frac \pi 2$ at the same value of $y$. In other words, they intersect. If we allow for topology change, such a surface can never be extremal. Nonetheless, such surfaces will exist when considering extremal surfaces homologous to half-spaces, which only have a single end point, i.e. a single corresponding extremal surface. Moreover, it seems that the problem can also be circumvented by displacing the end-points of the extremal surface of an internal in a time-like direction. In that case, the extremal surfaces do not intersect anymore in the singularity. Oddly, however, they cross each other at a finite distance while flipping the orientation of the entanglement wedge.



\section{The smallest extremal surface passes through the singularity}
\label{app:proof}
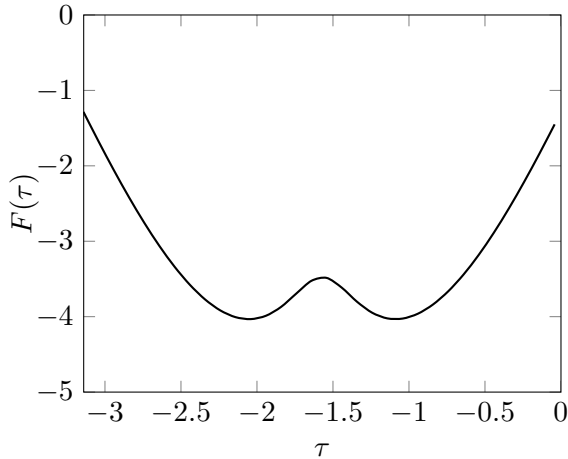
\begin{figure}
    \centering
    \begin{tikzpicture}
    \begin{axis}[xlabel=$\tau$,ylabel=$F(\tau)$, ylabel shift = -0.3cm, xmin=-3.14, xmax=0, ymin=-5, ymax=0, x=2cm, y=1cm]
    \addplot [smooth, thick] file {data/F.dat};
    \end{axis}
    \end{tikzpicture}
    \caption{The function $F$ of \cref{eq:func_F} for $y = -0.5$, $\alpha = 4$.}
    \label{fig:F}
\end{figure}
In this appendix, we will show that the extremal surfaces which one obtains by passing through the singularity indeed carry smaller areas. As argued in the main text, including such surfaces leads to a discontinuous phase transition in the entanglement entropy, which is physically forbidden.

For convenience, we will focus on the case $\alpha >0$, where there is some boundary time $\tau_*$ such that for $\tau_\bdry < \tau_*$ extremal surfaces can fall through the singularity. The crux of our argument will be to show that the last surface that falls into the singularity, i.e., the surface at $\tau_*$, at which the discontinuous phase transition happens, is the smallest extremal surface which ends on the boundary within $- \frac \pi 2 \leq \tau_\bdry \leq \frac \pi 2$.

By the RT formula, the RT surface for a boundary subregion $A$ is the smallest extremal surface in the bulk that ends on the boundary of $A$. Given two brane times $\tau_1$ and $\tau_2$ such that extremal surfaces emanating from $\tau_{1,2}$ intersect the asymptotic boundary at the same boundary time $\tau_\bdry$, it is straightforward to show that 
\begin{align}
    S^\text{(ren)}(\tau_2) > S^\text{(ren)}(\tau_1),
\end{align}
with $S^{(\text{ren})}$ discussed in the paragraph above \cref{eq:sint_full_2d}, can be rewritten as
\begin{align}
   F(\tau_2) > F(\tau_1),
\end{align}
with
\begin{align}
    \label{eq:func_F}
    F(\tau) = \alpha \sin(\tau) - \log \left(\tan y_\brane + \sqrt{1 + \mathcal Q_E(\tau) + \tan^2 y_\brane} \right).
\end{align}
\Cref{fig:F} shows a representative example of $F$ when the stronger bound \cref{eq:alpha_bound} is violated. Two properties are of interest to us. First, $F(\tau)$ is symmetric around $\tau = - \frac \pi 2$ and second, it has two minima at some $\tau_\text{min} = -\frac \pi 2 \pm \delta$. Clearly, the smallest extremal surfaces are the ones which intersect the brane at $\tau_\text{min}$.

The time $\tau_\text{min}$ can be compared to the brane time at which the last extremal surface passing through the singularity intersects the brane. This time is given by the local maximum of \cref{eq:rt_trajectory_2d_alt}. The crucial observation is that this maximum agrees with $\tau_\text{min}$. Thus, the last extremal surface to pass through the singularity is also the smallest one.

To demonstrate this observation we note that the extrema of \cref{eq:rt_trajectory_2d_alt} are located at
\begin{align}
    \label{eq:extrema}
    \alpha ^2  (z-1) \sin^4 \tau - 2 \alpha (z-1) \sin^3 \tau +\alpha ^2 (1-2 z) \sin^2 \tau +2 \alpha  \sin \tau z+\left(\alpha ^2+1\right) z = 0,
\end{align}
where we have defined
\begin{align}
    z = 1 + \alpha \cos^2 y_\brane.
\end{align}
\Cref{eq:extrema} is simply the derivative of \cref{eq:rt_trajectory_2d_alt} with respect to brane time, further multiplied by $\sqrt{1 + \alpha^2 \cos^2 \tau_\brane \cos^2 y_\brane}$. Similarly, setting the derivative of \cref{eq:func_F} multiplied by $\cos^2 y_\brane$ to zero we once again arrive at \cref{eq:extrema}. This proves our assertion.

\section{Computing extremal surfaces with varying endpoints}
\label{app:variation}
Finding the correct RT surfaces in the connected phase requires the use of the calculus of variations with varying endpoints, which we will review here for the convenience of the reader. The discussion will essentially focus on the type of situations encountered in this paper and point out differences between the Lorentzian and Euclidean case.

Let us start by considering the following integral,
\begin{align}
    I = \int_a^b L(x, y(x), y'(x)) dx,
\end{align}
the types of which we encountered when computing the RT surfaces. We now want to look for a stationary point without completely fixing the starting point $a$ of integration. Varying the integral and allowing for the variation of $a$ as well, it follows that
\begin{align}
    \label{eq:bdry_action_app}
    \delta I = \int_a^b dx \left( \frac{\delta L}{\delta y} - \frac{d}{dx} \frac{\delta L}{\delta y'} \right) \delta y - \frac{\delta L}{\delta y'} \delta y(a) - L(a, y(a), y'(a)) \delta a.
\end{align}
We now need to understand how $\delta y(a)$ is related to $\delta a$, if the initial point of $y(x)$, which we will denote by $y_i$, is constrained to lie on some surface. This means we have a constraint $y_i = f(a)$, where $f(x)$ describes the surface. Generally speaking, we have that
\begin{align}
    \delta y_i = (\delta y)(a) + y'(a) \delta a.
\end{align}
Using the constraint we also have that
\begin{align}
    \delta y_i = f'(a) \delta a.
\end{align}
Using these expressions, we can express the vanishing of the boundary term in \cref{eq:bdry_action_app} as
\begin{align}
    \label{eq:bdry_condition_vanilla_app}
    \frac{\delta L}{\delta y'} (y'(a) - f'(a)) - L(a, y(a), y'(a)) = 0.
\end{align}

In order to get some intuition for the result, let us take the case where we extremize a curve $y(x)$ which is supposed to end on a curve $f(x)$ in Euclidean flat space. In that case \cref{eq:bdry_condition_vanilla_app} becomes
\begin{align}
    y'(a) = -\frac 1 {f'(a)} \qquad (\text{for }\mathbb R^2).
\end{align}
This is nothing but the condition that an extremal curve which can end on a surface must end on that surface at a right angle.

The analogous case in two-dimensional Lorentzian space where $y$ is a time-like coordinate yields
\begin{align}
    y'(a) = \frac 1 {f'(a)} \qquad (\text{for }\mathbb R^{1,1})
\end{align}
which is the same orthogonality condition, now however in Lorentzian signature.

For the situations discussed in this paper, we encounter the case where the area functional contains a term which is evaluated at the brane. Such a term, which we will denote by $\phi(y_i)$ is varied according to
\begin{align}
    \delta \phi(y_i) = \partial_y \phi(y(a)) \delta y_i = \partial_y \phi(y(a)) f'(a) \delta a.
\end{align}
Thus, the corresponding boundary condition becomes 
\begin{align}
    \label{eq:bdry_condition_app}
    \frac{\delta L}{\delta y'} (y'(a) -f'(a)) - L(a, y(a), y'(a)) + \partial_y \phi(y(a)) f'(a) = 0.
\end{align}
Let us note here that had we chosen to vary the endpoint $b$, the sign of the last term would have been flipped. 

We can also, as above, evaluate this expression for the two-dimensional case and find
\begin{align}
     \frac{ y'(a)}{\sqrt{1 + y'(a)^2}}  = - \frac 1 {f'(a)\sqrt{1 + y'(a)^2}} + \partial_y \phi(y(a))& \qquad (\text{for }\mathbb R^2),\\
     \frac{ y'(a)}{\sqrt{1 - y'(a)^2}}  = + \frac 1 {f'(a)\sqrt{1 - y'(a)^2}} - \partial_y \phi(y(a))& \qquad (\text{for }\mathbb R^{1,1}).
\end{align}
The seemingly counter-intuitive point is that while in the Euclidean case the initial point $a$ of an extremal surface moves towards smaller values of $\phi$, it moves towards bigger values in the Lorentzian case.

\bibliographystyle{JHEP}
\bibliography{references}

\end{document}